%% file: alcor.tex
\PassOptionsToPackage{noend}{algorithmic}
\PassOptionsToPackage{dvipsnames}{xcolor}
\PassOptionsToPackage{breakspaces}{clevethm}
\PassOptionsToPackage{hypertexnames=false}{hyperref}
\PassOptionsToPackage{capitalize,nameinlink}{cleveref}
\PassOptionsToPackage{textsize=footnotesize}{todonotes}
\documentclass[journal,10pt,twoside]{IEEEtranTCOM}
\usepackage{graphics}
\usepackage{adjustbox}
\usepackage{pgfplots}
\usepackage[outdir=./images/]{epstopdf}
\usepackage{caption}
\usepackage{subcaption}
\usepackage{xspace}
\usepackage{cite}
\usepackage{float}
\usepackage{pgf}
\usepackage{tikz}
\usepackage{xcolor}
\usepackage[percent]{overpic}
\usepackage{textcomp}
\normalsize
\input{TeX/Preamble.tex}
\tikzexternaldisable

\title{A Deep Learning Based Resource Allocator for Communication Networks with Dynamic User Utility Demands}
\author{Pourya Behmandpoor, 
Mark Eisen,
Panagiotis Patrinos, 
Marc Moonen
\thanks{\funding \\\Affiliation}
}

\begin{document}
\maketitle
\copyrightnotice

    \input{TeX/Text/abstract.tex}

    \section{Introduction}\label{sec:intro}
        \input{TeX/Text/introduction.tex}

    \section{System Model}\label{sec:sys_model}
        \input{TeX/Text/system_model.tex}
    
    \section{Proposed Method}\label{sec:proposed_method}
        \subsection{Problem statement}\label{sec:problem}
            \input{TeX/Text/problemFormulation.tex}

        \subsection{Unconstrained resource allocation (URA)}\label{sec:URA}
            \input{TeX/Text/URA.tex}
        \subsection{Time-sharing}\label{sec:timesharing}
            \input{TeX/Text/timeSharing.tex}
        \subsection{Algorithm}\label{sec:algorithm}
            \input{TeX/Text/algorithm.tex}
        \subsection{Distributed ALCOR}\label{rem:distributed}
            \input{TeX/Text/distributedALCOR.tex}
        \subsection{Convergence study}\label{sec:conv}
            \input{TeX/Text/convergence.tex}
            
    \section{Numerical Experiments} \label{sec:sim}
        \input{TeX/Text/simulation.tex}
        \subsection{Circularly symmetric channel}\label{sec:sim:circ}

\input{TeX/Text/symm_channel.tex}

        \subsection{Multi-link per cell channel}\label{sec:sim:multilink}
            \input{TeX/Text/multi_channel.tex}
            
    \section{Conclusion and Future Directions}
        \input{TeX/Text/conclusion.tex}

\appendices
    \section{} \label{sec:aux:basics}
        \input{TeX/Text/appendix/notation.tex}

    \section{} \label{sec:aux:conv}
        \input{TeX/Text/appendix/conv.tex}

\bibliographystyle{IEEEtranTCOM}
\bibliography{IEEEabrv,TeX/bibliography.bib}


\end{document}

%% file: TeX/Preamble.tex
\usepackage{booktabs}
\usepackage{makecell}
\usepackage{float}
\usepackage{algorithmic}
\usepackage{bbm}
\usepackage{letltxmacro}
\usepackage[most]{tcolorbox}
\usepackage[%
	thmreset=section,
	eqreset=section,
	noload={algpseudocode,algorithmic}
]{myPreamble}

\pgfplotsset{compat=1.16}
\allowdisplaybreaks

\newlength\myindent
	
	\newcommand{\sqnrm}[1]{\left\Vert#1\right\Vert^2}
	\newcommand{\nrm}[1]{\left\Vert#1\right\Vert}

\makeatletter
	\newcommand{\E}{\@ifstar\@@E\@E}					
	\newcommandx{\@E}[3][1={},3={}]{
	\mathbb E_{#1}\ifstrempty{#2}{}{
		\left[
			#2\ifstrempty{#3}{}{\mid #3}
		\right]
	}
	}
	\newcommandx{\@@E}[3][1=k,3={}]{
	\mathbb E_{#1}\ifstrempty{#2}{}{
		[#2\ifstrempty{#3}{}{|#3}]
	}
	}
\makeatother

\Crefname{ALC@unique}{Line}{Lines}
\renewcommand{\algfont}{\bf}

\newlist{algsubstates}{enumerate}{2}
\makeatletter
	\setlist[algsubstates,1]{
		label={\alph*:},
		ref={\theALC@line.\alph*},
		itemsep=0pt,
		partopsep=0pt,
		topsep=0pt,
		parsep=0pt,
	}
\makeatother
\newcommand{\fixstate}{\addtocounter{ALC@line}{-1}\refstepcounter{ALC@line}\phantomsection}
\newcommand{\State}{\STATE\fixstate}

\makeatletter
	\renewcommand\theALC@line{{\oldstylenums\arabic{ALC@line}}}
	\def\theHALC@line{\thealgorithm-\arabic{ALC@line}}
\makeatother
\crefname{ALC@line}{step}{steps}
\Crefname{ALC@line}{Step}{Steps}
\crefalias{algsubstatesi}{ALC@line}

\newlist{tblsubstates}{enumerate}{2}
\makeatletter
	\setlist[tblsubstates,1]{
		label={\alph*:},
		ref={\theALC@line.\alph*},
		itemsep=0pt,
		partopsep=0pt,
		topsep=0pt,
		parsep=0pt,
	}
\makeatother
\crefname{TBL@line}{table}{tabless}
\Crefname{TBL@line}{table}{Tabless}
\crefalias{tblsubstatesi}{TBL@line}

\newenvironment{talign}
 {\let\displaystyle\textstyle\align}
 {\endalign}

\newenvironment{talign*}
 {\let\displaystyle\textstyle\csname align*\endcsname}
 {\endalign}

 \newcommand{\new}[1]{#1}

 \newcommand{\Affiliation}{%
 Pourya Behmandpoor, Panagiotis Patrinos, and Marc Moonen are with KU Leuven University, Department of Electrical Engineering (ESAT), STADIUS Center for Dynamical Systems, Signal Processing and Data Analytics (e-mail: pourya.behmandpoor, marc.moonen, panos.patrinos @esat.kuleuven.be).
 Mark Eisen is with Johns Hopkins University, Applied Physics Laboratory (e-mail: mark.eisen@ieee.org).
 }

 \newcommand{\funding}{This research work was carried out at the ESAT Laboratory of KU Leuven, in the frame of Research Project FWO nr. G0C0623N 'User-centric distributed signal processing algorithms for next generation cell-free massive MIMO based wireless communication networks' and Fonds de la Recherche Scientifique - FNRS and Fonds voor Wetenschappelijk Onderzoek - Vlaanderen EOS Project no 30452698 '(MUSE-WINET) MUlti-SErvice WIreless NETworks'. 
 The Work is also supported by the FWO research projects G081222N, G033822N, and G0A0920N; Research Council KU Leuven C1 project No. C14/24/103;
 The scientific responsibility is assumed by its authors.
 }

\newcommand\copyrighttext{%
  \footnotesize \textcopyright\ 2025 IEEE.  Personal use of this material is permitted.  Permission from IEEE must be obtained for all other uses, in any current or future media, including reprinting/republishing this material for advertising or promotional purposes, creating new collective works, for resale or redistribution to servers or lists, or reuse of any copyrighted component of this work in other works.}
  
\newcommand\copyrightnotice{%
\begin{tikzpicture}[remember picture,overlay]
\node[anchor=south,yshift=5pt] at (current page.south) {\fbox{\parbox{\dimexpr\textwidth-\fboxsep-\fboxrule\relax}{\copyrighttext}}};
\end{tikzpicture}%
}

%% file: TeX/Text/abstract.tex
\begin{abstract}
    Deep learning (DL) based resource allocation (RA) has recently gained significant attention due to its performance efficiency. However, most related studies assume an ideal case where the number of users and their utility demands, \eg data rate constraints, are fixed, and the designed DL-based RA scheme exploits a policy trained only for these fixed parameters. Consequently, computationally complex policy retraining is required whenever these parameters change.
    In this paper, we introduce a DL-based resource allocator (ALCOR) that allows users to adjust their utility demands freely, such as based on their application layer requirements. {ALCOR employs deep neural networks (DNNs) as the policy in a time-sharing problem.} The underlying optimization algorithm iteratively optimizes the on-off status of users to satisfy their utility demands in expectation. The policy performs unconstrained RA (URA)--—RA without considering user utility demands--—among active users to maximize the sum utility (SU) at each time instant. {Depending on the chosen URA scheme, ALCOR can perform RA in either a centralized or distributed scenario.}
    The derived convergence analyses provide theoretical guarantees for ALCOR's convergence, and numerical experiments corroborate its effectiveness compared to meta-learning and reinforcement learning approaches.
\end{abstract}

\begin{keywords}
    Deep learning based resource allocation, dynamic data rate constraints, dynamic quality-of-service, utility demands, centralized and distributed resource allocation
\end{keywords}

%% file: TeX/Text/introduction.tex
Resource allocation (RA) in communication systems has been an active research topic for decades \cite{mhatre_interference_2007,cendrillon_autonomous_2007,chiang_power_2007,shi_iteratively_2011}, resulting in the development of various RA schemes. Since the underlying optimization problem is nonconvex and possibly large-scale, conventional RA schemes typically exhibit slow convergence in large communication systems. Moreover, these schemes cannot be easily extended either to consider longer time horizons, \eg to perform RA for an average of utilities over time, or to address the vast interconnection of users taking their various mutual effects into account. These limitations motivate recent advancements in deep learning (DL)-based RA, where a policy, such as a deep neural network (DNN), is typically dedicated to translating dynamic parameters of the communication system, \eg channel coefficients, into optimal resources, \eg the optimal transmit power for each user \cite{zhou_dynamic_2019,tong_nine_2022}.

Existing DL-based RA schemes in the literature incorporate various training strategies. The policies are either trained in a supervised manner \cite{sun_learning_2018}, where RA solutions are available as labels, or in an unsupervised manner \cite{liang_towards_2020,lee_resource_2018,liang_deep-learning-based_2020,cui_spatial_2019}, where a utility function, \eg sum data rate of users, is considered the global reward function.
Alternatively, policies can serve as decision-makers that monitor the state space, \eg the set of channel coefficients, and choose actions within the action space, \eg different transmit power levels. Such policies can be trained using reinforcement learning (RL) principles \cite{liang_spectrum_2019,ye_deep_2019,guo_reinforcement_2022,nasir_multi-agent_2019}. 
The mentioned training strategies are mostly model-based; they assume a model for the reward, and the trainer optimizes the policy parameters using the calculated gradient of this model.
An alternative training approach is model-free, where the gradient is approximated by various methods, such as policy gradient (REINFORCE \cite{sutton1999policy}) or zeroth-order optimization, by measuring reward values \cite{eisen_learning_2019,kalogerias_model-free_2020,wang_learning_2022}. This training can also be performed in real-time while the communication system is operating. Through real-time reward measurement, this approach can capture the full behavior of the system, including nonidealities such as nonlinearities in the modulator and demodulator or antenna setup, which are not typically captured by model-based approaches due to simplifications.

In the inference step, existing DL-based RA schemes are employed in either a \emph{centralized} or a \emph{distributed} manner.
In a centralized approach, a server is responsible for conducting the RA by gathering necessary information from all users and employing a centralized policy for RA \cite{ye_deep_2019,nasir_multi-agent_2019,liang_spectrum_2019,wang_learning_2022}.
{Distributed RA can serve as an alternative to centralized RA, enabling users to locally decide how to allocate shared communication resources without transmitting relevant information, such as channel state information, to a central server. Distributed RA is particularly beneficial in scenarios where a server with sufficient communication and computational capacity is unavailable within the network and where avoiding a single point of failure is critical. In this approach, message-passing occurs locally between neighboring users who interfere with each other the most, ensuring more efficient use of communication bandwidth \cite{ji2023meta}, scalability (\eg by using graph neural networks (GNNs) \cite{wu_comprehensive_2021,shen_graph_2021,wang_learning_2022,naderializadeh_learning_2022,eisen_optimal_2020,naderializadeh_state-augmented_2022,huang2024meta}), and a reduction in deployment delays and RA policy complexity \cite{liang_spectrum_2019, nasir_multi-agent_2019, ye_deep_2019,nguyen2021drl,tian2021multiagent}.
These benefits come at the cost of limited local memory and computational resources, as well as more complex procedures to maintain the necessary synchrony between users during both training and inference steps. 
In fully connected networks, where each user can directly exchange messages with all other users, ensuring synchrony and identifying neighboring groups is more straightforward compared to networks with limited connectivity between users.}
In the distributed setting, the policies can be either identical among the users \cite{ye_deep_2019,nasir_multi-agent_2019} or different \cite{liang_spectrum_2019,behmandpoor2024asynchronous}. Furthermore, the training step can also be classified as either centralized or distributed \cite{ye_deep_2019,nasir_multi-agent_2019,liang_spectrum_2019,wang_learning_2022,behmandpoor_MODEL-FREE_2022,huang2024meta,tian2021multiagent,behmandpoor_federated_2022}.

DL-based RA can be categorized into \emph{unconstrained} RA and \emph{constrained} RA. In unconstrained RA, a global reward function is optimized with minimal constraints, such as simple box constraints that can be met by applying a projection through an appropriate output activation function. However, this approach cannot address more complex constraints, such as user utility (quality-of-service (QoS)) demands—e.g., user data rate constraints. This limitation means the RA is unable to allocate additional resources to users with higher demands. A simple workaround involves optimizing the weighted sum of utilities as the global reward function \cite{behmandpoor_deep_2021, cui_spatial_2019}, where the weights reflect users' utility demands. Conversely, in constrained RA, a global reward function is optimized while satisfying more complex constraints. During policy training, these constraints are typically managed either by including a penalization term in the reward function \cite{liang_towards_2020, peng2019deep, tian2021multiagent, wu2020dynamic}, or by using the primal-dual optimization method, which directly addresses constrained optimization problems \cite{kalogerias_model-free_2020, naderializadeh_learning_2022, wang_learning_2022, eisen_optimal_2020, huang2024meta}.

Constrained DL-based RA schemes in the literature utilize policies, such as DNNs, which are trained for a fixed set of RA constraints (e.g., utility demands). 
{For example, in works such as \cite{tian2021multiagent,ye_deep_2019,peng2019deep,wu2020dynamic}, RA, service migration, and radio access network slicing are addressed using RL for vehicular communication, where delay-sensitive and delay-tolerant, as RA constraints, are considered through a penalization term in the reward function. Although these works consider heterogeneous QoS and other RA constraints in their optimization problem to guide the policy toward meeting these constraints in dynamic RA, their trained policies cannot generalize to a new set of constraints—for example, when a user increases their data rate demand.}
In such RA schemes, policy retraining becomes necessary if the RA constraints change. 
This limitation undermines the applicability of existing DL-based RA schemes, as policy retraining is computationally complex, leading to delays in real-time deployment.
One possible workaround is to incorporate dynamic constraints, such as utility demands, along with all other dynamic components of the communication system, into the policy input. In this case, the policy would be trained for various utility demands. However, generalizing the problem with a moderate amount of resources, such as the number of layers and neurons in DNN policies, is not practically feasible. This has already been experimentally validated in \cite{cui_spatial_2019} and our simulations.
{This limitation has recently sparked research into developing resilient policies with the capability to generalize to new network conditions, such as new RA constraints.
Works such as \cite{dong_deep_2021} in bandwidth and transmit power allocation, and \cite{yuan2020transfer} in beamforming,} utilize transfer learning concepts \cite{pan_survey_2010} to fine-tune only part of the policy (e.g., the last few layers) to reduce the computational complexity of retraining.
{Furthermore, many studies such as \cite{yuan2022adapting,qu2021dmro,wang2020fast,huang2020meta} in DL-based resource scheduling and task offloading, and \cite{ji2023meta,huang2024meta,yuan2021meta} in DL-based RA, leverage meta-learning concepts \cite{finn2017model,vettoruzzo2024advances,nagabandi2018learning,li2016learning} to initialize a policy that can quickly adapt to varying network conditions with only a few training samples. In \cite{ji2023meta}, a distributed RA method is proposed to maximize user energy efficiency while meeting their minimum SNR requirements to guarantee utility (QoS) demands. The RA policy is trained based on RL in a federated learning framework to allocate optimal subchannels and transmit power. Constraints are incorporated into the reward function by penalizing decisions that violate them, similar to the approach in \cite{ye_deep_2019,yuan2021meta} for V2V communication. Additionally, distributed transmit power allocation is investigated in \cite{huang2024meta}, where GNNs are used to locally determine the optimal transmit power to increase the sum rate (SR). In line with \cite{naderializadeh_learning_2022}, data rate constraints are satisfied (on average over time, as in our setting) by penalizing violations while following primal-dual updates.
In all the aforementioned methods \cite{yuan2022adapting, qu2021dmro, wang2020fast, huang2020meta, ji2023meta, huang2024meta, yuan2021meta}, constraint violations are penalized during training. Additionally, these methods leverage model-agnostic meta-learning (MAML) principles \cite{finn2017model}, where they first train an initial policy on a subset of constraints, and then adapt to new constraints (e.g., utility demands) by fine-tuning the entire policy with a few additional training samples.}

{Although meta-learning has shown promising results in rapidly adapting policies to new network conditions, it still involves policy fine-tuning, which may be computationally intensive for complex policies, such as large DNNs. Moreover, the initial policy training in meta-learning requires a more computationally demanding training procedure. The generalizability of the policy after fine-tuning also depends on the diversity and representativeness of the training samples, as well as the similarity of nework conditions \cite{finn2017model,hospedales2021meta,nguyen2021drl}, which cannot always be easily guaranteed in RA.
Among various network conditions, in this paper, we focus on user utility demands, allowing users to have dynamic utility demands (such as dynamic user data rate constraints) based on their real-time tasks.
Existing DL-based RA methods, as well as DL-based resource scheduling and task offloading methods, require policy re-training or fine-tuning to address a new set of user utility demands. To the best of our knowledge, the proposed work is the first DL-based RA method that adapts to dynamic user utility demands without any policy retraining or fine-tuning.}

\bigskip
\noindent
\emph{Objective:} 
{In this paper\footnote{This paper is an extension of the work presented in \cite{behmandpoor_learning-based_2022}.}, the objective is to propose a DL-based RA scheme that can adapt, on average over time, to dynamic user utility demands.
The novelty of this work lies in the fact that, unlike existing DL-based RA schemes, the adaptation to new sets of user utility demands does not require any retraining {or fine-tuning} of the involved policies, thereby conserving computational resources and improving speed.}

\noindent
\emph{Contributions:} 
The contributions are listed as follows:
\begin{enumerate}
    \item {A RA scheme is proposed, which involves time-sharing and DL-based RA among the users.
    At each time instant, the proposed RA scheme selects a subset of users to be activated and performs DL-based RA among these activated users to maximize their sum utility (SU). The proposed time-sharing algorithm is an optimization algorithm that iteratively controls the on-off status of users to guarantee their utility demands in expectation.}
    \item {Convergence analyses are provided for the proposed RA scheme under standard assumptions, deriving a convergence rate of $\mathcal{O}(\nicefrac{1}{\sqrt{k}})$, with $k$ representing the iteration counter in the considered time-sharing algorithm.}
    \item {Rigorous numerical experiments are conducted to assess the performance of the proposed RA scheme and its distributed variant against various benchmarks.
    }
\end{enumerate}

%% file: TeX/Text/system_model.tex
We consider $N$ users (links) each equipped with a transmitter and a receiver. 
The direct channel between the transmitter and receiver of user $i$ is denoted by $h_{ii}$, while the interference channel between the transmitter of user $j$ and the receiver of user $i$ is denoted by $h_{ij}$. All the channel coefficients define the full channel matrix $\bm{H} \in \mathbb{C}^{N \times N}$ with $h_{ij}$ as its element in the $i$th row and $j$th column. 

For the RA, we consider a central deep neural network (DNN) as the policy defined by
\begin{align}
    \func{\phi(&\bm S, \bm{\theta})}{\R^{s} \times \bm \Theta}{\R^{\new{r}}} 
    \label{eq:new_policy}
\end{align}
with input $\bm S$ and policy parameter $\bm{\theta} \in \bm\Theta \subset \R^n$.
Here $\bm S$ is a random variable incorporating global or local measurements of users, which vary over time and is a function of parameters such as the global channel $\bm{H}$ and the users' state. Examples of elements in $\bm S$ could include channel coefficients, packet queue length awaiting transmission, or received interference power from neighbors (refer to \cref{sec:sim} for more details). 
The policy parameter $\bm{\theta}$ may contain DNN weights and biases, to be optimized during training.
The policy output specifies resources (ideally to be optimal) such as transmit power, frequency band, timeslot, beamformer angle, etc. of all $N$ users, \new{with an $r$-dimensional output}. Furthermore, each user $i$ has a utility function $U_i(\bm{H},\phi(\bm{S},\bm{\theta}))$, which depends on the communication channel $\bm{H}$ and the resources allocated to all users by $\phi(\bm{S},\bm{\theta})$. For instance, if the utility represents the user data rate, $U_i = R_i(\bm{H}, \bm{p})$, it depends on the channel matrix $\bm{H}$ and the transmit power of all users $\bm{p} = (p_1,\dots,p_N)$ (due to interference between users), as set by the centralized policy output, i.e., $\bm{p} = \phi(\bm{S},\bm{\theta})$. Refer to \cref{sec:sim} and \eqref{eq:data_rate} for further details.
It is noted that the policy described in \eqref{eq:new_policy} is designed for centralized RA, where a single server performs RA. Distributed extensions will be discussed in \cref{rem:distributed}.

%% file: TeX/Text/problemFormulation.tex
\input{TeX/Tab/notations.tex}

Before formally stating the problem, consider the user selection vector $\bm\xi$ defined as follows:
\begin{defin}[user selection vector (USV) $\bm\xi$]\label{def:TSV}
    Define the USV as $\bm \xi = (\xi_1,\dots,\xi_N) \in \Omega^\xi \coloneqq \{0, 1\}^N$. 
    The elements $\xi_i$ are independently drawn from a Bernoulli distribution \cite{edition2002probability} with mean: 
    \begin{equation}
        \kappa_i := \mathbb{E}\{\xi_i\} \in [0,1], \quad \forall i \in [N].
        \label{eq:kappa}
    \end{equation}
    By the USV, user $i$ is selected to be active if $\xi_i = 1$, otherwise it is switched off.
    Moreover, the USV Bernoulli distribution is denoted by $\mathcal{D}^\xi(\bm\kappa)$ with $\bm\kappa \coloneqq (\kappa_1,\cdots,\kappa_N) \in [0,1]^N$.
\end{defin}
\noindent
Note that according to \eqref{eq:kappa}, the probability that user $i$ is activated is $\kappa_i$. 
{Moreover, the Bernoulli distribution, characterized by its binary variables, provides a straightforward and interpretable definition. Its first moment directly corresponds to the user activation probability and can be easily implemented by the server (in the centralized variant) or by individual users (in the distributed variant) by flipping a biased coin}.

After defining the USV $\bm\xi$, we are interested in the following problems:
\newline

\noindent
\textbf{Time-sharing problem:}\\
Find the optimal probabilities $\bm\kappa = (\kappa_1,\cdots,\kappa_N) \in {[0,1]^N}$ for activating users at each time instant, based on $\bm\xi \sim \mathcal{D}^\xi(\bm\kappa)$, such that:
\begin{itemize}
    \item The utility demands of all users are satisfied in expectation {(on average over time)};
\end{itemize}
\vspace{6pt}
\noindent
\textbf{Resource allocation problem:}\\
At each time instant, allocate resources among the users such that:
\begin{itemize}
    \item The SU is maximized among active users for a given $\bm\xi\in\Omega^\xi$ and $\bm{H}$.
\end{itemize}
\noindent
After optimizing the probabilities $\bm\kappa$, at each time instant, a random instance of the USV $\bm\xi$ defined in \cref{def:TSV} is drawn.
Then, according to $\bm\xi$, some users are activated while others are deactivated. Subsequently, a URA is performed among the active users. 

Before addressing the mentioned problems, the basic assumptions made throughout the paper are as follows:
\begin{ass}[basic assumptions]\label{ass:basic}
    \begin{enumerate}
        \item \label{ass:basic:2} {The dynamic parameters of the communication network, \eg the channel $\bm{H}$, are ergodic stochastic variables \cite{edition2002probability}, with possibly unknown distributions;}
        \item \label{ass:basic:3} The utility demands $\bm{u}^{\min}$ are changing slower than the dynamic parameters of the communication network, and are assumed to be constant in time windows.
    \end{enumerate}
\end{ass}
\noindent
{It is remarked that in \cref{ass:basic}, utility demands $\bm{u}^{\min}$ are not assumed to be fixed, which is a restrictive assumption in the literature that we aim to relax. Moreover, \cref{ass:basic:2} regarding the dynamic parameters of the communication network is a standard assumption in the literature \cite{ye_deep_2019,nasir_multi-agent_2019,eisen_optimal_2020,naderializadeh_state-augmented_2022}. \cref{ass:basic:3} is a mild assumption in practice since utility demands $\bm{u}^{\min}$ mostly reflect the demands originating from the application layer, \eg following a task scheduling scheme based on currently running tasks. These demands are slower in nature, typically changing at a rate of seconds or more, compared to the channel dynamics in wireless communication networks, which typically change at a rate of milliseconds in fast-fading scenarios (refer to \cref{sec:algorithm} for further explanation).}

%% file: TeX/Tab/notations.tex
\begin{table*}[t]
    \centering
    \begin{tabular}{c c | c c}
        \toprule
        not. & Description & not. & Description\\
        \midrule
        $N$ & number of users & $\phi(\cdot, \bm\theta)$ & parameterized policy\\
        $\bm H$ & channel matrix & $U_i^\theta(\bm H_{\xi})$ & utility of user $i$\\
        $\bm X_\xi$ & $\bm X$ for activated users & $r_i$ & resource allocated to user $i$\\
        $\bm \theta$ & policy parameter & $u_i^{\min}$ & utility demand of user $i$\\
        $\bm \kappa$ & activation probabilities of users & $\hat{F}_i(\bm H_\xi)$ & $u_i^{\min} - U_i^{\theta}(\bm H_\xi)$\\
        $\bm \xi$ & user selection vector ($\kappa_i = \E[\xi]{\xi_i}$) & $F_i(\bm \lambda)$ & $\E[H,\xi]{\hat{F}_i(\bm H_\xi)}$\\
        $\bm \lambda$ & decision vector, determining $\bm\kappa$ &  $B$ & batch size\\
        \bottomrule
    \end{tabular}
    \caption{List of notations}
    \label{tab:notations}
\end{table*}

%% file: TeX/Text/URA.tex
{To address the DL-based URA for each $\bm\xi$ and $\bm{H}$, the RA policy $\phi$ defined in \eqref{eq:new_policy} is trained using the following unconstrained optimization:}
\begin{align*}
    {\maximize_{\bm{\theta}\in\R^n}}&{\sum_{i\in[N]}\E[\bm{H}\sim\mathcal{D}^H, \bm\kappa\sim\mathcal{D}^{\kappa}, \xi\sim\mathcal{D}^\xi(\bm\kappa)]{U^{\theta}_i(\bm{H}_\xi)}}, \label{eq:policyTraining} \numberthis\\
    \text{where}~~ &U^{\theta}_i(\bm{H}_\xi) \coloneqq U_i(\bm{H}_\xi,\phi(\bm S_\xi,\bm{\theta})),
\end{align*}
with $U_i(\bm{H}_\xi,\phi(\bm S_\xi,\bm{\theta}))$ as the utility of user $i$, explained in \cref{sec:sys_model}. 
Here, $\bm{H}_{\xi}$ denotes the global channel for the subset of active users, which equals the channel matrix $\bm{H}$ with distribution $\mathcal{D}^H$ except for rows and columns corresponding to the zero elements of $\bm \xi$ (deactivated users), which are set to zero. 
Similarly, $\bm S_\xi$ contains measurements of the active users, and $\bm\kappa \coloneqq (\kappa_1,\cdots,\kappa_N)$ with a uniform distribution $\mathcal{D}^\kappa$ in $[0,1]^N$. For simplicity, the policy input $\bm S_\xi$ is assumed to be a function of only the channel $\bm{H}$, so is $U^{\theta}_i$. 

In problem \eqref{eq:policyTraining}, the policy parameter is optimized to maximize the SU of users in expectation, serving as a global utility function. This problem can be addressed using first-order stochastic gradient descent (SGD) over batches of random samples.  
This leads to the following iterations with an initial value $\bm{\theta}^0$ and an iteration counter $\ell\in[L-1]$ with $L > 0$:
\begin{align}
    \bm{\theta}^{\ell+1} = \bm{\theta}^\ell + \frac{\gamma^\ell}{B} \sum_{i\in[N]} \sum_{j=1}^{B}{\nabla_{\bm{\theta}} \bm{U}^{\bm{\theta}^\ell}_i(\bm{H}_\xi^{\ell,j})} \label{eq:training}.
\end{align}
In \eqref{eq:training}, sample averaging is performed over batches of size $B$ with channels $\bm{H}_\xi^{\ell,j}$ and probabilities $\bm\kappa^{\ell,j}$ where $\bm\xi^{\ell,j} \sim \mathcal{D}^\xi(\bm\kappa^{\ell,j})$. Here, the superscript $\{\ell,j\}$ indicates the $j$th sample in the batch taken at iteration $\ell$. The step size is also denoted by $\gamma^\ell$. 
The capability of policy $\phi$ to generalize the URA problem for all possible $\bm{H}, \bm\xi$, and $\bm\kappa$ heavily depends on the policy resources, such as the number of layers and perceptrons in each layer in the case of a DNN policy. 
Due to the limited memory and computational capacities, as well as the nonconvex nature of the problem, generalization is often suboptimal in practice.

The policy $\phi$ and the training procedure used to address the maximization problem in \eqref{eq:policyTraining} depend on the URA method employed. For instance, existing URA methods in the literature, including those utilizing various training procedures other than \eqref{eq:training}, such as RL, can be incorporated. The proposed RA scheme is agnostic to the choice of the URA method (refer to \cref{rem:blackboxURA}).

During the inference stage, the \emph{trained} policy is utilized to perform URA among active users at each time instant as follows:
\begin{align*}
    r_i \coloneqq [\phi(\bm S_\xi, {\bm{\theta}^L})]_i \quad &\text{if} \quad \xi_i=1, \label{eq:URA} \numberthis\\
    r_i \coloneqq 0 \quad &\text{if} \quad \xi_i = 0,
\end{align*}
where $[\cdot]_i$ indicates the $i$th element of the vector and $r_i$ defines the allocated resource for user $i$.
The frequently used notations are summarized in \cref{tab:notations}.

\new{
It is remarked that the term \emph{unconstrained} in URA emphasizes that dynamic user utility demands—such as varying data rate or latency requirements—are not addressed in this stage of resource allocation. However, minimal constraints, such as simple box constraints, can still be handled within the URA. For example, a maximum transmit power limit can be enforced by applying a sigmoid activation function in the final layer of $\phi$.
}

%% file: TeX/Text/timeSharing.tex
After training the policy $\phi(\cdot,\bm{\theta}^L)$ for URA, we proceed to formulate the time-sharing problem as defined in \cref{sec:problem}.
To do so, we first define the following function $\func{F}{\R^N}{\R^N}$
\begin{align*}
    F(\bm\lambda) \coloneqq  \E[\bm{H}, \bm\xi\sim \mathcal{D}^\xi(\bm\kappa)]{\hat F(\bm{H}_\xi)}, & \numberthis \label{eq:F}\\
    \text{where}\quad
    \hat F(\bm{H}_\xi) \coloneqq&~ \bm u^{\min} - \bm{U}^{\bm{\theta}^L}(\bm{H}_\xi),\\
    \kappa_i =&~ \nicefrac{(1+\lambda_i)}{\max_\ell\{1+\lambda_\ell\}},
\end{align*}
with $\bm\lambda \in \R^N_+$, $\bm\kappa$ defined in \cref{def:TSV}, and $\bm{U}^{\bm{\theta}^L} \coloneqq (U_1^{\bm{\theta}^L},\cdots,U_N^{\bm{\theta}^L})$.

{In \eqref{eq:F}, we aim to solve the following problem:
\begin{equation}
    \text{find} ~\bm\lambda\in\R^N_+ ~~\text{such that}~~ F(\bm\lambda) \leq \bm 0,
    \label{eq:timesharing}
\end{equation}
where $\bm 0 \in \R^N$ is a vector of all zeros and $\leq$ is an element-wise operator.
The equality $[F(\bm\lambda)]_i=0$ indicates that the expected utility of user $i$, $\E[\bm{H}, \bm\xi\sim \mathcal{D}^\xi(\bm\kappa)]{U_i^{\bm{\theta}^L}(\bm{H}_\xi)}$, is equal to its utility demand $\bm{u}^{\min}_i$, while $[F(\bm\lambda)]_i < 0$ indicates that the expected utility of user $i$ is greater than its demand. 
To guarantee inequality \eqref{eq:timesharing}, we propose using $\bm\lambda$ to calculate the probabilities $\kappa_i$. }
This choice is motivated by the fact that $\kappa_i$ remains nonzero for any choice of $\bm\lambda$ in \eqref{eq:timesharing}, maintaining higher utility for the corresponding user, and yet, whenever required, $\kappa_i$ can become arbitrarily small by choosing large $\lambda_j$ by other users $j$. 
Thus, the coupling between users, which is due to interference in the communication system, can be captured by this definition of $\kappa_i$. 

To estimate the expectation in \eqref{eq:F}, we define the following sample average:
\begin{align}
    \textstyle \hat{\bar{F}}(\mathbb{H}_{\bm\xi}^{k}) \coloneqq \frac{1}{B} \sum_{j=1}^{B} \hat F(\bm{H}_\xi^{k,j}),
    \label{eq:meanF}
\end{align}
where $\mathbb{H}_{\bm\xi}^{k}$ refers to the batch of samples, defined as $\mathbb{H}_{\bm\xi}^{k} \coloneqq \{\bm{H}_\xi^{k,1},\cdots,\bm{H}_\xi^{k,B}\}$ and $\bm{H}_\xi^{k,j}$ represents the $j$th sample in a batch of size $B$, taken at iteration $k$ with USV $\bm\xi^k$.
The algorithm for finding $\bm\lambda$ in \eqref{eq:timesharing} is outlined in the next subsection.

%% file: TeX/Text/algorithm.tex
{The proposed algorithm for the time-sharing problem \eqref{eq:timesharing} is summarized in \cref{alg:centralized}, which can be easily implemented using the mappings defined in \eqref{eq:F} and \eqref{eq:meanF} with a batch of channel samples $\bar{\mathbb{H}}_{\bm\xi}^{k} \coloneqq \{\bar{\bm{H}}_\xi^{k,1},\cdots,\bar{\bm{H}}_\xi^{k,B}\}$.
\begin{algorithm}[t]
    \input{TeX/Alg/algCentralized_pro.tex}
    \caption{ALCOR}
    \label{alg:centralized}
\end{algorithm}
At each iteration $k$, in \cref{alg:step:1}, the algorithm activates a subset of users according to the USV instance $\bar{\bm\xi}^{k,j} \sim \mathcal{D}^\xi(\bar{\bm\kappa}^k)$, which depends on the most updated probabilities $\bar{\bm\kappa}^k \coloneqq (\bar\kappa_1^k,\cdots,\bar\kappa_N^k)$. Then, in \cref{alg:step:1.5}, URA is performed among the activated users to maximize their SU. This procedure is repeated $B$ times for each $j\in[B]$ to form a batch, allowing the calculation of the sample average \eqref{eq:meanF} with the samples $\bar{\mathbb{H}}_{\bm\xi}^{k}$.
Following \cref{alg:step:2}, the current parameter $\bm\lambda^k$ is updated using either a fixed stepsize $\alpha_k=\alpha$ or a diminishing stepsize $\alpha_k=\nicefrac{\alpha_0}{\sqrt{1+\tilde{\alpha}k}}$ with some $\alpha_0, \tilde{\alpha} > 0$, where \Cref{alg:step:3} ensures the nonnegativity of updates, \ie $\bm\lambda^k \in \R^N_+$.
The same steps are repeated for a new batch of samples $\mathbb{H}_{\bm\xi}^{k}$ with $\bm\xi^{k,j} \sim \mathcal{D}^\xi(\bm\kappa^k)$, based on the probabilities $\bm\kappa^k$ set in \cref{alg:step:6}. Updates are performed following \cref{alg:step:5,alg:step:5.5}.
At the end, in \cref{alg:step:7}, the probabilities $\bar{\bm\kappa}^{k+1}$ are updated using the parameters $\bar{\bm\lambda}^{k+1}$.
Note that in the implementation, $\bar\kappa_i^{k+1}=0$ may be assigned whenever $\bar\kappa_i^{k+1} < 0$, which can occur due to negative values of $\bar\lambda_i^{k+1}$ during the initial iterations before the algorithm has converged.}
During the iterations, the utility demands are assumed to remain fixed within a time window of length $T$, which is sufficiently long to allow the algorithm to meet the utility demands on average over time, {\ie $K \ll T$. 
Moreover, once utility demands are updated, the algorithm restarts with a new input $\bm{u}^{\min}$.}

{It is noted that the two-step updates (\cref{alg:step:3,alg:step:5}) can be seen as extragradient-like updates \cite{hsieh2020explore,pethick2023solving}, where during one iteration, the algorithm first generates an intermediate iterate $\bm{h}^{k}$ from a base iterate $\bar{\bm\lambda}^k$ and then completes the update by taking a step from the base iterate using the intermediate iterate. The time-sharing problem \eqref{eq:timesharing} can also be implemented by a simpler algorithm where a batch of samples \(\bm\xi^{k,j} \sim \mathcal{D}^\xi(\bm\kappa^k)\) is employed, followed by URA among the activated users and calculating the average in \eqref{eq:meanF}. 
Then the parameter $\bm{\lambda}^k$ can be updated by 
\begin{equation*}
    \bm{\lambda}^{k+1} = \max\left\{\bm 0, \bm{\lambda}^k + \alpha_k \hat{\bar{F}}({\mathbb{H}}_{\bm\xi}^k)\right\}
\end{equation*}
at each iteration $k$. 
This update rule guarantees an increase in $\lambda_i$, and subsequently the probability $\kappa_i$, whenever the average utility of user $i$ is below its demand $\bm{u}^{\min}_i$.
However, there is no convergence guarantee for this simpler update rule in the optimization literature.} Convergence analysis for \cref{alg:centralized} is provided in the Appendix, where the time-sharing problem \eqref{eq:timesharing} is cast as an inclusion problem under the class of nonmonotone variational inequalities \cite{diakonikolas2021efficient}.

\begin{rem}[{agnostic to URA methods}]\label{rem:blackboxURA}
    The proposed RA scheme can employ policies as a black box for URA. These policies can be centralized or distributed and can be trained and utilized by various DL methods, such as RL, GNN, etc. \cite{liang_towards_2020,shen_graph_2021,ye_deep_2019}.
    However, the scalability of the proposed method, the maximum number of users that ALCOR can accommodate for RA, and the communication overhead in the distributed variant depend on the employed URA method. 
    Hence, suitable methods need to be selected based on the communication system of interest. 
    \new{Refer to \cref{sec:sim:multilink} for numerical studies on ALCOR's performance with various URA methods.}
\end{rem}

ALCOR is limited to scenarios that can be formulated by \eqref{eq:F} and \eqref{eq:timesharing}. Specifically, ALCOR can address RA problems where increasing $\kappa_i$, the probability of user $i$ being activated, leads to an increase in the user's utility and consequently satisfies its demand.

%% file: TeX/Alg/algCentralized_pro.tex
\begin{algorithmic}[1]
    \item[\algfont{Input}]
    {utility demands $\bm{u}^{\min}$,} stepsizes $\gamma > 0, ~{\alpha_k} \in (0,1)$,
    \item[] \quad~\(\text{batch size } B > 0\)
    
    \item[\algfont{Initialize}]
    \(\bar{\bm\lambda}^{-1}=\bar{\bm\lambda}^{0} \in \R^N,~ \bm{h}^{0} \in \R^N, {~\bar{\bm\kappa}^0 \in \R^N}\)
        
    \hspace{5pt}
    \item[\algfont{Repeat} $k=0,1,\dots,K-1$]
    \State\label{alg:step:1}
    Sample a batch of \(\bar{\bm\xi}^{k,j} \sim \mathcal{D}^\xi(\bar{\bm\kappa}^k)\) with $j\in[B]$%
    \State\label{alg:step:1.5}
    Perform URA \eqref{eq:URA} and form the average $\hat{\bar{F}}(\bar{\mathbb{H}}_{\bm\xi}^k)$ in \eqref{eq:meanF}%
    \State\label{alg:step:2}
    \(\bm{h}^{k} \coloneqq \bar{\bm\lambda}^k + \gamma \hat{\bar{F}}(\bar{\mathbb{H}}_{\bm\xi}^k) + (1 - \alpha_k)\left(\bm{h}^{k-1} - \bar{\bm\lambda}^{k-1} - \gamma \hat{\bar{F}}(\bar{\mathbb{H}}_{\bm\xi}^k)\right)\)%
    \State\label{alg:step:3}\label{alg:step:barz}
    \(\bm\lambda^{k} = \max\{\bm 0, \bm{h}^{k}\}\)%
    \State\label{alg:step:6}   
    \(\kappa_i^{k} = \nicefrac{(1+\lambda_i^{k})}{\max_{\ell}\{1+\lambda_\ell^{k}\}}, \quad \forall i\in [N]\)%
    \State\label{alg:step:4}
    Sample a batch of \(\bm\xi^{k,j} \sim \mathcal{D}^\xi(\bm\kappa^k)\) with $j\in[B]$%
    \State\label{alg:step:5.5}
    {Perform URA \eqref{eq:URA} and form the average $\hat{\bar{F}}(\mathbb{H}_{\bm\xi}^k)$ in \eqref{eq:meanF}}%
    \State\label{alg:step:5}\label{alg:step:lambda_update}
    \(\bar{\bm\lambda}^{k+1} = \bar{\bm\lambda}^k - \alpha_k \left(\bm{h}^{k} - \bm\lambda^{k} - \gamma\hat{\bar{F}}(\mathbb{H}_{\bm\xi}^k)\right)\)%
    \State\label{alg:step:7}   
    \(\bar\kappa_i^{k+1} = \nicefrac{(1+\bar\lambda_i^{k+1})}{\max_{\ell}\{1+\bar\lambda_\ell^{k+1}\}}, \quad \forall i\in [N]\)%

\end{algorithmic}

%% file: TeX/Text/distributedALCOR.tex
{Distributed RA can serve as an alternative to centralized RA, enabling users to locally decide how to allocate shared communication resources, where a server with sufficient communication and computational capacity is unavailable within the network. Refer to \cref{sec:intro} for further details on the benefits and limitations of distributed RA.}

{
ALCOR can be readily extended to a distributed variant by employing distributed URA methods.
In the distributed variant, each user has an individual policy $\phi_i(\cdot, \bm\theta_i)$ to perform distributed URA (see for example \cite{liang_spectrum_2019, nasir_multi-agent_2019, ye_deep_2019}). In this case, the policies are trained according to \eqref{eq:policyTraining}, with $U^{\theta}_i(\bm{H}_\xi) \coloneqq U_i(\bm{H}_\xi,\bm\phi(\bm S_\xi,\bm{\theta}))$, where $\bm\phi(\bm S_\xi,\bm{\theta}) \coloneqq (\phi_1(\bm S_{\xi,1},\bm\theta_1), \cdots, \phi_N(\bm S_{\xi,N},\bm\theta_N))$, $\bm\theta = (\bm\theta_1,...,\bm\theta_N)$, and $\bm S_\xi = (\bm S_{\xi,1},...,\bm S_{\xi,N})$. Similar to the centralized case, due to the interference between users, the utility $U^{\theta}_i$ depends on the allocated resources of all users. However, in the distributed scenario, resources are defined by individual policies $\phi_i(\bm S_{\xi,i},\bm\theta_i)$ rather than a centralized policy $\phi(\bm S_{\xi},\bm\theta)$. The individual policies consider individual local measurements $\bm S_{\xi,i}$ and their parameters may be different, i.e., $\bm\theta_i \neq \bm\theta_j$ for $i \neq j$ \cite{liang_spectrum_2019}, or there may be a consensus among them, i.e., $\bm\theta_i = \bm\theta_j$ for $i \neq j$ \cite{nasir_multi-agent_2019,ye_deep_2019}, depending on the employed distributed URA scheme.
URA in \eqref{eq:URA} is performed with $r_i = \phi_i(\bm S_{\xi,i}, \bm{\theta}^L_i)$ if $\xi_i = 1$, and $r_i = 0$ otherwise. It is noteworthy that in distributed ALCOR, although the utility of user $i$, $U^{\theta}_i$, depends on all individual policies, user $i$ only measures $U^{\theta}_i$ without requiring other local parameters $\bm\theta_j, j\neq i$. 
Considering the definition of the mapping $\hat F$ in \eqref{eq:F}, each user $i$ locally executes \cref{alg:centralized} with local variables $\lambda_i^k, \bar\lambda_i^k, h_i^k$, and $\kappa_i^k$, utilizing local estimates $[\hat{\bar{F}}(\mathbb{H}_{\bm\xi}^k)]_i$ and $[\hat{\bar{F}}(\bar{\mathbb{H}}_{\bm\xi}^k)]_i$, where $[\cdot]_i$ indicates the $i$th element of the input vector.
Refer to \cref{sec:sim} for an example.}

{
Since ALCOR performs URA to allocate resources among activated users at each time instant, its performance depends on the employed URA method. Consequently, estimation errors in the URA policy input values can also affect overall performance, particularly in distributed scenarios, which are more prone to such errors due to the limited local computational capacities of users. To address this common issue in existing DL-based RA methods in the literature, \cite{cui2023uncertainty} proposes an uncertainty injection algorithm during training, which enhances the generalizability of the policy against input estimation errors. As ALCOR is agnostic to the choice of URA method, incorporating such robust URA methods is feasible. Further investigation to improve URA generalizability within the framework of ALCOR can serve as a direction for future research.
}

{
To update $\kappa_i^k$ in \cref{alg:step:6}, each user requires the normalization term $\max_{\ell}\{1+\lambda_\ell^{k}\}$, which can be obtained by exchanging scalar values $\lambda_\ell^{k}$ among users via message-passing. This incurs an additional communication overhead in the distributed scenario. It is worth noting that obtaining local measurements $\bm S_{\xi,i}$ (see \cref{sec:sim}, the \emph{distributed policy} paragraph, for an example of local measurements) also requires message-passing, which occurs at each time instant. In contrast, the scalar values $\lambda_\ell^{k}$ are communicated only after batches of samples ($2B$ time instants). Therefore, the processing latency and communication overhead incurred solely by the distributed ALCOR method are negligible compared to those of the employed distributed URA method \cite{liang_spectrum_2019, nasir_multi-agent_2019, ye_deep_2019,nguyen2021drl,tian2021multiagent,ji2023meta,wang_learning_2022}.
\new{Refer to \cref{sec:sim} for numerical communication overhead comparisons.}
}

%% file: TeX/Text/convergence.tex
{
The convergence rate of \cref{alg:centralized} is provided in the following theorem. The formal convergence statement and its proof are presented in Appendix \ref{sec:aux:conv}.
\begin{thm}[informal]
    \cref{alg:centralized} converges with a rate of $\mathcal{O}(1/\sqrt{k})$, where $k$ denotes the iteration number.
\end{thm}
\noindent
This theorem states that for a sufficiently large time window length $T$, ALCOR converges to an optimal time-sharing policy within the scope of the time window, and users can operate while meeting their current utility demands.
}

%% file: TeX/Text/simulation.tex
In this section, we evaluate the proposed RA scheme through numerical experiments across various communication scenarios. We address the power allocation problem where user utilities are defined as data rates. 
{Specifically, the resources are the transmit powers $\bm{r} = \bm{p} = (p_1,\cdots,p_N) \in [0,p^{\max}]^N$, constrained by a maximum $p^{\max}$, and utilities are data rates, where $U_i = R_i$ and}
\begin{equation}
     R_i(\bm{H},\bm{p}) \coloneqq \log_2{\left(1+\frac{|h_{ii}|^2p_i}{\sigma_n^2+\sum_{j \neq i}{|h_{ij}|^2p_j}}\right)},
     \label{eq:data_rate}
\end{equation}
with $\sigma_n^2$ as the power of independent and identically distributed (IID) additive white Gaussian noise at the receivers. 
Two centralized and distributed URA schemes, adopted from \cite{liang_towards_2020,nasir_multi-agent_2019}, are employed for policy training and performing URA among the activated users.

\bigskip
\noindent \textbf{Centralized policy:} A fully connected DNN is considered as the URA policy, following the structure outlined in  \cite{liang_towards_2020}. {Specifically, it consists of $4$ layers with the number of neurons set to $\{400, 400, 200, 20\}$, unless specified otherwise. }
The policy input is the full channel matrix $\bm{H}$, denoted as $\bm S=\bm{H}$ in \eqref{eq:new_policy}, and the output is the continuous transmit power of all users $\bm p$. 
{With the considered policy, URA can be performed for a maximum number of 20 users, although the number of active users per time instant may be fewer.}
The activation function of hidden layers is the rectified linear unit (relu), while for the output layer, the \emph{sigmoid} function is considered to ensure compliance with the transmit power box constraint. Batch normalization is also applied during training in all layers except the output layer.
The policy structure is the same throughout the simulations unless specified otherwise. 

\bigskip
\noindent \textbf{Distributed policy:} 
A fully connected deep neural network (DNN) is considered as the URA policy, comprising 4 layers with the number of neurons set to {$\{41, 100, 50, 1\}$}. As shown in \cite{nasir_multi-agent_2019}, all users utilize the same DNN, i.e., $\phi_1=\dots=\phi_N$. Each user collects local measurements to feed into its local policy and adjusts its continuous transmit power based on the policy output. The activation functions are consistent with the centralized case, and batch normalization is also employed.
Define the following sets
\begin{align*}
    \mathcal{I}_i^t &\coloneqq \{j \in [N], j \neq i \mid |h_{ij}^t|^2 p_j^t > \eta \sigma^2\},\\
    \mathcal{O}_i^t &\coloneqq \{j \in [N], j \neq i \mid |h_{ji}^t|^2 p_i^t > \eta \sigma^2\},
\end{align*}
for user $i$ at time $t$. Here, $\mathcal{I}_i^t$ represents the set of users at time $t$ causing interference to user $i$, where the interference power exceeds the threshold $\eta \sigma^2$. Conversely, $\mathcal{O}_i^t$ represents the set of users at time $t$ receiving interference from user $i$ exceeding the same threshold.
{In the simulations, the maximum cardinality of these sets is limited to 5 by selecting 5 most affected users in $\mathcal{O}_i^t$ and 5 most affecting users in $\mathcal{I}_i^t$, and the parameter $\eta$ is fixed at $1$.}
The local measurements of user $i$ at time $t$, used as input for its policy $\phi_i$, are obtained from \cite{nasir_multi-agent_2019} and are included here for completeness:
\begin{itemize}
    [leftmargin=0.5cm,
    ]
    \item Transmit power $p_i^{t-1}$ and data rate $R_i^{t-1}$;
    \item Direct channels $|h_{ii}^t|^2$ and $|h^{t-1}_{ii}|^2$;
    \item Received interference power from all users in two time instants: $\sum_{j=1, j \neq i}^N |h_{ij}^{t}|^2 p_j^{t-1} + \sigma^2, \sum_{j=1, j\neq i}^N |h_{ij}^{t-1}|^2 p_j^{t-2} + \sigma^2$;
    \item Received interference power from interfering neighbors in the set $\mathcal{I}_i$ in two timeslots: $\{|h_{ij}^{t}|^2 p_j^{t-1} \mid j \in \mathcal{I}_i^{t-1}\}, \{|h_{ij}^{t-1}|^2 p_j^{t-2} \mid j \in \mathcal{I}_i^{t-2}\}$;
    \item Data rate of the interfering neighbors: $\{R_j^{t-1} \mid j \in \mathcal{I}_i^{t-1}\}, \{R_j^{t-2} \mid j \in \mathcal{I}_i^{t-2}\}$;
    \item Normalized transmitted interference power to affected neighbors in the set $\mathcal{O}_i$ as $\{|h_{ji}^{t'_i}|^2 p_i^{t'_i} \left(\sum_{i=1, i \neq j}^N |h_{ji}^{t-1}|^2 p_i^{t-1} + \sigma^2\right)^{-1} \mid j \in \mathcal{O}_i^{t'_i}\}$, where $t'_i$ is the last time when user $i$ was active;
    \item Direct channel of interfered neighbors: $\{|h_{jj}^{t-1}|^2 \mid j \in \mathcal{O}_i^{t'_i}\}$;
    \item Data rate of interfered neighbors: $\{R_j^{t-1} \mid j \in \mathcal{O}_i^{t'_i}\}$.
\end{itemize}
{The input set $\bm S_{i}$ for each user $i$ comprises $41$ elements, forming the local policy input. 
If the sets $\mathcal{I}_i$ and $\mathcal{O}_i$ contain fewer than $5$ elements, zero-padding is applied at the respective policy input locations to ensure a fixed input size.}
The policy uses past measurements to accommodate channel instances that are correlated over time. 
However, initial simulations suggest that this correlation has negligible impact on the proposed RA scheme performance on the considered tasks. Therefore, we report the results with IID channel samples.

\begin{figure}[t]
    \centering
    \begin{overpic}[width=0.8\columnwidth]{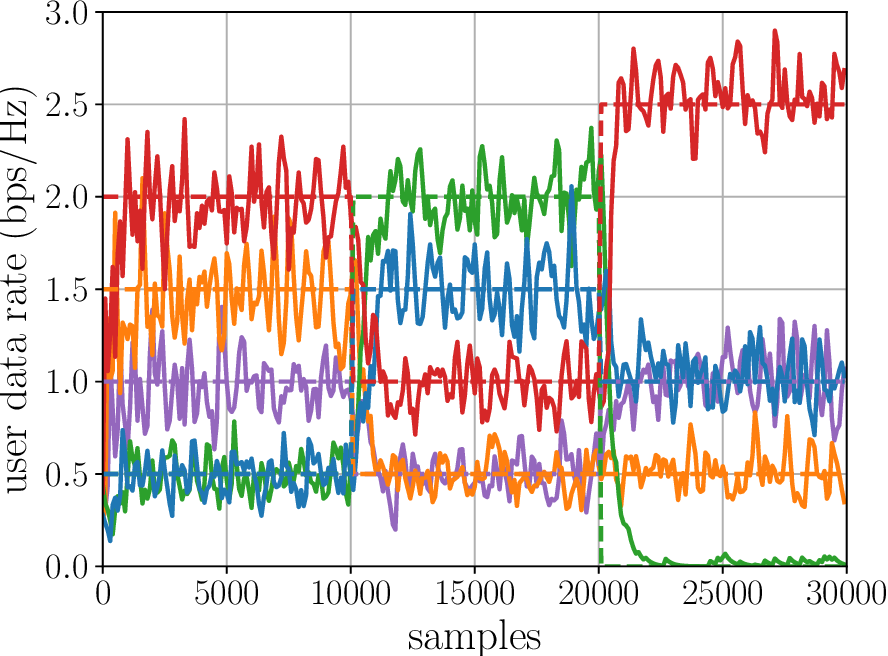}
    \end{overpic}
    \caption{
        {Satisfying utility demands (on average) in a $5$-user centralized scenario using ALCOR. Three time windows with different utility demands, $\bm u^{\rm min}$, are shown. Each solid line represents the instantaneous data rate of a user over time, and the dashed lines represent the dynamic utility demands of the corresponding colors.}}
    \label{fig:changing_rates}
\end{figure}

\bigskip
\noindent \textbf{Policy training:} 
In both centralized and distributed scenarios, policies are trained using an unsupervised approach, as discussed in \cite{liang_towards_2020}, where the reward function during the training is \eqref{eq:policyTraining}. The distributions for $\bm{H}, \bm\kappa$, and $\bm\xi$ vary in different scenarios and are specified in the sequel.

\bigskip
\noindent \textbf{Benchmarks:}
\begin{enumerate}
    [leftmargin=0.5cm,
    ]
    \item \textbf{Fixed DNN:} 
    A DNN is considered in the centralized scenario for each user number $N$, where the utility demands $\bm u^{\rm fix} \in \R^N$ are fixed and are to be satisfied in expectation. {The detailed architecture of the DNNs will be mentioned subsequently for each $N$.}
    The DNNs are trained in an unsupervised manner \cite{liang_towards_2020} using the reward function
    {
    \begin{talign}
        \begin{split}
            \rm{reward}&(\bm{\theta}, \bm u^{\rm fix}) = \sum_{i=1}^N \E[\bm{H}]{U^{\theta}_i(\bm{H})} \\
            &- L \sum_{i=1}^N \max \{0, u_i^{\rm fix} - \E[\bm{H}]{U^{\theta}_i(\bm{H})}\},
        \end{split}
        \label{eq:dnn_fix}
    \end{talign}
    where constraint violations are penalized with $L=100$.
    During training, expectations are approximated using batches of size $1000$.} Note that this reward function differs from the one in \cite[eq. (14)]{liang_towards_2020}, where the DNN is forced to satisfy the utility demands for every channel instance $\bm{H}$.  

    
    \item \textbf{WMMSE:} The conventional iterative URA scheme WMMSE \cite{shi_iteratively_2011} is considered for assessing the trained policies in URA and as a black box URA scheme used within the proposed RA scheme.
    
    \item \textbf{GP:} The iterative geometric programming (GP) RA scheme \cite{chiang_power_2007} is also considered, which is able to satisfy the utility demands for each channel instance. When comparing with this method, only channel instances feasible for the GP optimization problem are considered in the comparison.
    
    \item \textbf{Meta-learning:} 
    {
        Existing works in the literature, such as DL-based RA methods in \cite{ji2023meta,huang2024meta,yuan2021meta}, and DL-based resource scheduling and task offloading methods in \cite{yuan2022adapting,qu2021dmro,wang2020fast,huang2020meta}, adapt to new network conditions (\eg new user utility demands) by utilizing principles of meta-learning (refer to \cref{sec:intro} for a literature review). Motivated by the success of meta-learning in these works, we consider this learning framework as another benchmark.
        In these works, constraint violations are penalized in the reward, and the MAML \cite{finn2017model} principles are employed to initialize their policy, enabling quick adaptation to a new set of utility demands. For our specific RA problem, and in line with the mentioned works, we define the reward function as the one in \eqref{eq:dnn_fix} and initialize the policy using $\bm{\theta}$ optimized by:}
    \begin{align}
        \begin{split}
            &\maximize_{\bm{\theta}\in\R^n, \psi>0} ~ \E[\bm{u}^{\min}]{\rm{reward}(\bm{\theta}^+, \bm{u}^{\min})}\\
            &\stt ~ \bm{\theta}^+ \coloneqq \bm{\theta} + \psi \nabla_{\bm{\theta}} \rm{reward}(\bm{\theta},\bm{u}^{\min}).
        \end{split}
        \label{eq:meta_learning}
    \end{align}
    {where we use the same policy structure for a fair comparison, and $\bm{u}^{\min}$ is uniformly sampled from the interval $[0,2.5]^N$.} In the simulations, the expectation in \eqref{eq:meta_learning} is approximated using exponential averaging with a decay factor of $0.7$ over batches of size 1000 during training and 50 during inference. The optimization parameters $\bm{\theta}$ and $\psi$ in \eqref{eq:meta_learning} are optimized following the procedure outlined in \cite{nagabandi2018learning} with step sizes of $10^{-4}$ and $5\times10^{-6}$, respectively.
\new{
    \item \textbf{Reinforcement Learning:}
    RL \cite{ji2023meta,tian2021multiagent,ye_deep_2019} is considered to address RA with dynamic user utility demands. Relevant simulation specifications are provided in \cref{sec:sim:multilink}.
}
\end{enumerate}

\begin{figure*}[th]
    \centering
    \begin{subfigure}[b]{0.33\textwidth}
        \begin{overpic}[width=\textwidth]{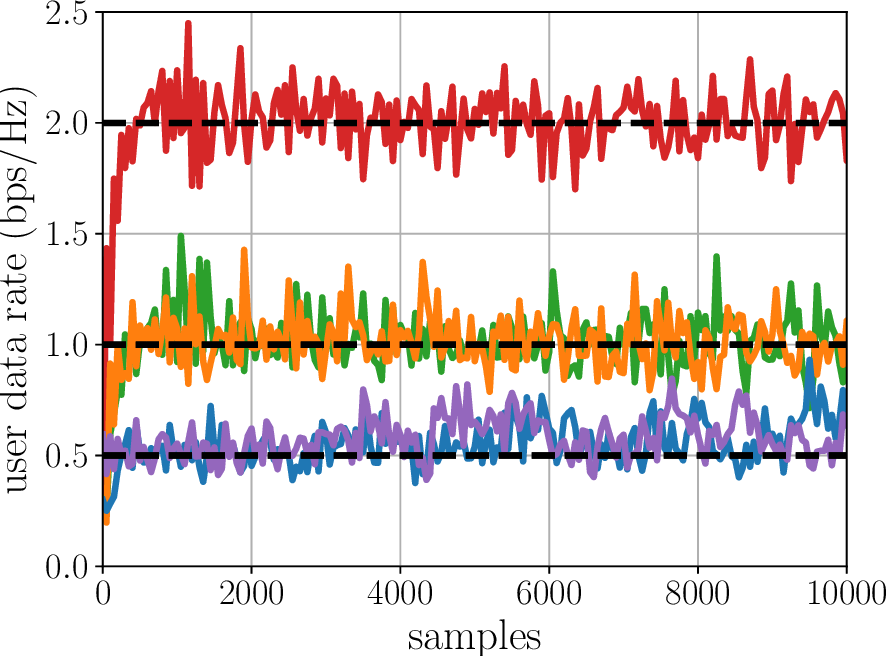}
        \end{overpic}
        \caption{{ALCOR}}
    \end{subfigure}%
    \begin{subfigure}[b]{0.33\textwidth}
        \begin{overpic}[width=\textwidth]{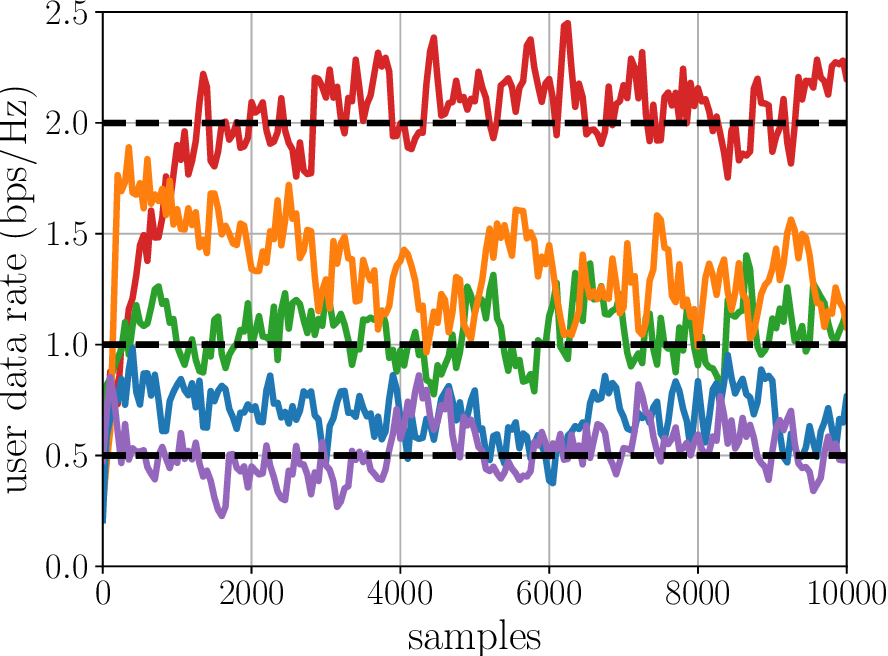}
        \end{overpic}
        \caption{{Meta-Learning}}
    \end{subfigure}%
    \begin{subfigure}[b]{0.33\textwidth}
        \begin{overpic}[width=\textwidth]{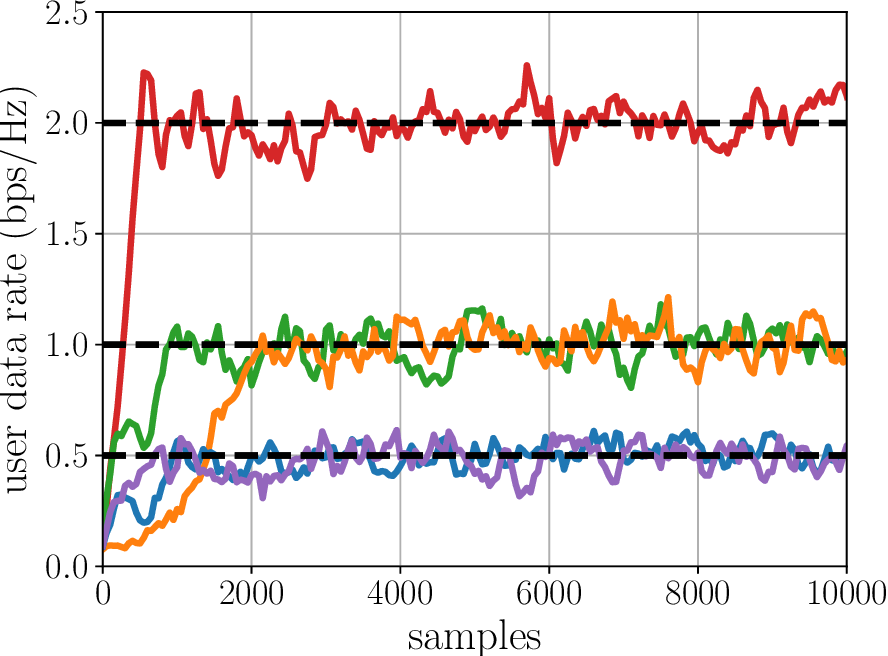}
        \end{overpic}
        \caption{{Vanilla Learning}}
    \end{subfigure}
    \caption{
        {Performance comparison of different RA schemes in a 5-user centralized scenario with a similar DNN structure of $\{25,25,25,5\}$. {\textbf{Each color specifies the instantaneous data rate of a user over time}}, where the utility demands (data rate demands) $\bm u^{\rm min}=(0.5,0.5,1,1,2)$ are depicted by black dashed lines. While the proposed scheme (ALCOR) does not involve any DNN retraining, meta-learning and vanilla learning (re)train the DNN using the reward function of \eqref{eq:dnn_fix}. The (re)training is performed with a diminishing step size of $\nicefrac{\psi^\star}{1+k}$, where $k$ is the epoch counter and $\psi^\star \approx 0.3$ is the optimized decision variable of \eqref{eq:meta_learning}. ALCOR demonstrates faster and more stable convergence.}
    }
    \label{fig:comparison}
\end{figure*}

\bigskip
{In the simulations, the step sizes of \cref{alg:centralized} are set as constants, with $\alpha_k=0.9$ and suitable values of $\gamma$ ranging from 1 to 10, depending on the specific experiment. Unless specified otherwise, the batch size is set to $B=25$, {and the utility demands $\bm u^{\min}$ are chosen such that the presentation is clear and the feasibility is guaranteed.}
In the rest of this section, two different channel scenarios are considered for performance evaluation.}

%% file: TeX/Text/symm_channel.tex
In this section, the channel coefficients are assumed to follow a circularly symmetric complex normal distribution, i.e. $h_{ij} \sim \mathcal{CN}(0,1)$, {$p^{\max}=1$}, and the SNR is set to $15$ dB.

\bigskip
\noindent
{\textbf{Adaptivity:}} 
In the first experiment, ALCOR's adaptivity is illustrated in \cref{fig:changing_rates}. In a scenario with 5 users, the utility demands values are set as $\bm{u}^{\min} = (0.5, 0.5, 1, 1.5, 2)$, $\bm{u}^{\min} = (2, 1.5, 0.5, 0.5, 1)$, and $\bm{u}^{\min} = (0, 1, 1, 0.5, 2.5)$ in $bps/Hz$ over three consecutive time windows. {For clarity of presentation, the chosen utility demands are strict, ensuring that the average user data rates match these demands after convergence.} The figure clearly shows that the proposed RA scheme can accurately adapt to the changing utility demands and meet them on average within a few iterations after each utility demand update. \new{In the sequel, we focus on a single time window—starting from an update in utility demands—to study ALCOR's performance and compare it with other RA methods.}

The performance of the proposed RA scheme in addressing dynamic utility demands is compared with the \emph{vanilla learning} and meta-learning approaches. In both approaches, the RA policy is retrained in an unsupervised manner using the reward function in \eqref{eq:dnn_fix} for the new utility demands. In vanilla learning, the policy is initialized randomly, while in meta-learning, the policy is initialized using the optimization in \eqref{eq:meta_learning}, with a single update for $\bm{\theta}^+$ \cite{nagabandi2018learning}.
It should be noted that meta-learning requires more complex policy training following \eqref{eq:meta_learning}, which involves second-order (Hessian) optimization, as considered here, or first-order optimization, as in \eg \cite{huang2024meta}. According to \cref{fig:comparison}, ALCOR adapts to user utility demands with simpler updates following \cref{alg:centralized}, \ie without fine-tuning the policy. Its adaptation is faster and more stable compared to the other approaches.
Although the objective of the considered meta-learning approach is to find a policy capable of adapting to new utility demands with a single update—\ie initializing the policy with the optimal $\bm{\theta}$ and then fine-tuning it with the optimal step size $\psi$ optimized by \eqref{eq:meta_learning}—it is evident that meta-learning cannot adapt to utility demands with a single update. Therefore, in our experiments, meta-learning adaptation continues with additional updates with a diminishing step size, in line with existing works on RA, \eg \cite{ji2023meta,huang2024meta,yuan2021meta}.
Compared to vanilla learning, meta-learning satisfies utility demands more quickly due to its educated initialization. However, its slower adaptation compared to ALCOR can be attributed to the fact that RA tasks (with different utility demands) share limited high-level structures, necessitating more iterations for fine-tuning the policy.
The complexity involved in meta-learning—both in training (initializing) the policy via \eqref{eq:meta_learning} and in fine-tuning it when addressing new sets of utility demands—underscore the motivation for using ALCOR in the considered constrained DL-based RA task.

\bigskip
\noindent
{\textbf{Unconstrained RA:}} In the next experiment, we assess the performance of the centralized and distributed policies, which are trained using \eqref{eq:policyTraining}, for URA as compared to WMMSE. 
Two training strategies are employed: 
1) \emph{Diverse training}: During training, $\bm\kappa$ is randomly and uniformly chosen from the set $\bm\kappa \in \{0.2 \times \bm 1, 0.5 \times \bm 1, \bm 1\}$, where $\bm 1$ is a vector of all ones. At each time instant, each user is independently switched on with a probability of $\kappa_i$ during the training. 
2) \emph{Non-diverse training}: During training, $\bm\kappa=\bm 1$, \ie all users are always switched on.

The performance of the policies in URA is presented in \cref{tab:URA}.
Based on \cref{tab:URA}, diverse training results in a slightly better generalization compared to non-diverse training. The performance of the policies in URA is also comparable to WMMSE, \ie the policies maintain good performance in URA while making URA faster. 
The same performance is evident in \cite{liang_towards_2020}. 
As diverse and non-diverse training strategies are comparable in performance, for the remaining experiments, we employ non-diverse training, which is the standard training method in the URA literature.

\input{TeX/Tab/ura.tex}
\input{TeX/Tab/circ_gp.tex}
\input{TeX/Tab/circ_tight.tex}

\bigskip
\noindent
{\textbf{Meeting utility demands:}} In the next experiment, the proposed RA scheme is compared against the GP method and the DNNs trained using the reward function \eqref{eq:dnn_fix} for  fixed utility demands (fixed DNNs). As the GP method requires a feasible RA problem at each time instant, only channels feasible for the corresponding utility demands are considered. The feasibility is determined using the procedure outlined in \cite{qian_mapel_2009}.
Centralized and distributed policies are trained using both feasible and infeasible training samples with performance assessed under varying numbers of users in the communication system. 
{The fixed DNNs have the following structures $\{100,400,200,10\}$, $\{400,400,200,20\}$, and $\{2500,400,400,50\}$ for $10$, $20$, and $50$ users, respectively. 
\cref{tab:circ:GP} reports the achieved SR of users while the utility demands are satisfied, on average. 
The user utility demands are reported in the caption where we use the notation $u^{\rm min}_{a:b}$ to refer to the utility demand of users $i \in \{a,a+1,\cdots,b\}$.}
The results indicate that the proposed method outperforms the GP method in terms of SR. Furthermore, although the fixed DNNs, trained for fixed utility demands, achieve the highest SR, they can only satisfy the fixed utility demands for which they were trained. A notable advantage of the proposed method is that it is not limited to feasible channel instances, unlike methods such as the one proposed in \cite{liang_towards_2020}. Consequently, the proposed method does not require the computationally complex feasibility check during training.
The utility demand violations are not reported in \cref{tab:circ:GP}, as they are equal to zero for all the reported RA schemes.

In the following, ALCOR is assessed under stricter utility demands. 
In this scenario, the problem is infeasible for most test samples, and the GP method is unable to provide any solution. The utility demand violation is reported as:
\begin{equation}
    \text{viol.} \coloneqq \max_{i \in [N]} \{\max\{0, u^{\rm min}_i -  R_i\}/{u^{\rm min}_i} \} \times 100.
    \label{eq:loss}
\end{equation}
Additionally, the fixed DNNs are the same as those in \cref{tab:circ:GP}. The results are summarized in \cref{tab:circ:tight}. According to the results, the proposed RA scheme can accurately satisfy dynamic utility demands while maintaining a high SR. As expected, the fixed DNNs cannot meet the new set of utility demands, and they require retraining of their policies. 
\new{The users' average data rates are also plotted in \cref{fig:circ:tight} for the distributed case with $N=20$ (see the supplementary material for additional plots).}
\begin{figure}[t]
    \centering
    \begin{subfigure}[b]{\columnwidth}
        \includegraphics[trim=0em 0em 0em 0em, clip, width=\columnwidth]{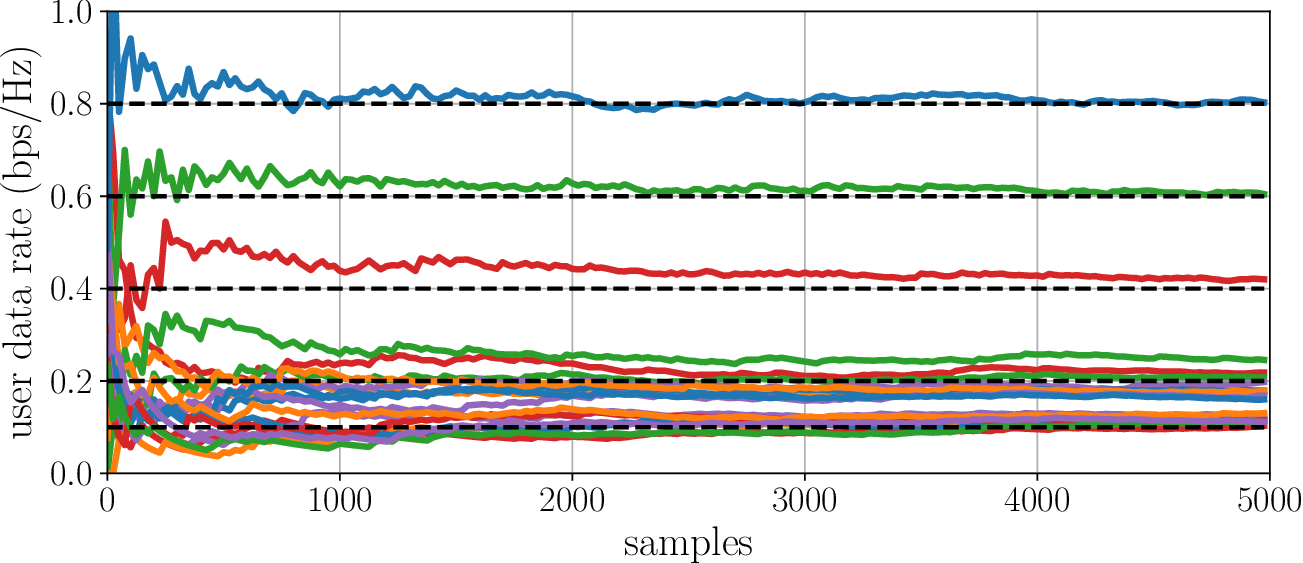}
        \caption{\new{Moving average of data rates. Each line represents a user's data rate moving average, and the dashed lines indicate the corresponding utility demands.}}
    \end{subfigure}
    \begin{subfigure}[b]{\columnwidth}
        \includegraphics[trim=0em 0em 0em 14em, clip, width=\columnwidth]{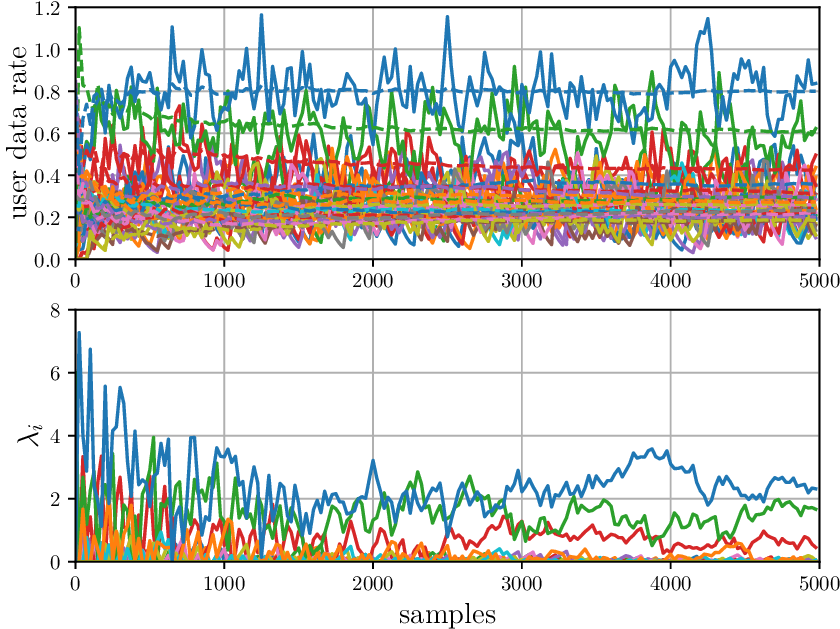}
        \caption{\new{$\lambda_i$ values corresponding to the data rates in (a), with matching colors.}}
    \end{subfigure}
    \caption{
        \new{User utilities and the corresponding $\lambda_i$ values generated by ALCOR for the distributed RA scenario reported in \cref{tab:circ:tight}, with $N=20$.}      
        }
    \label{fig:circ:tight}
\end{figure}
It is evident that the user data rates rapidly converge to meet the demands upon any change. The values $\bm\lambda$ are also plotted, showing higher values for users with greater demands. This finding aligns with what is expected from \eqref{eq:F} and \eqref{eq:timesharing}, where higher values of $\lambda_i$ increase the probability $\kappa_i$ of user $i$ being active, thereby pushing its expected utility toward higher values.

\input{TeX/Tab/variousURA.tex}
\new{Regarding the communication overhead incurred during the distributed RA, the URA policy adopted from \cite{nasir_multi-agent_2019} requires sharing 25 variables per channel update per user. Considering a channel coherence time of 10 ms and 32-bit floating-point precision per value, the communication overhead (without compression) is roughly 78 kbps per user. To obtain the normalization terms in \cref{alg:step:6}, ALCOR needs to share (or update upon any change) the normalization term $\max_{\ell}\{1+\lambda_\ell^{k}\}$ every $2B$ time instants among the users. Our experiments show an average rate of 67 bps per user, which is negligible compared to the overhead incurred by the employed distributed URA and other existing RA methods \cite{liang_spectrum_2019, nasir_multi-agent_2019, ye_deep_2019,nguyen2021drl,tian2021multiagent,ji2023meta,wang_learning_2022}.}



%% file: TeX/Tab/ura.tex
\begin{table*}[th]
    \centering
    \begin{tabular}{c c c c c c}
        \toprule
        $\bm\kappa$ & cntr. (divs.) & cntr. (non-divs.) & dist. (divs.) & dist. (non-divs.) & WMMSE\\
        \midrule
        0.25 & 5.88 & 5.63 & 5.82 & 5.45 & 5.88\\
        0.5 & 6.38 & 6.30 & 6.31 & 6.09 & 6.8\\
        0.75 & 6.56 & 6.56 & 6.52 & 6.35 & 7.29\\
        1 & 6.71 & 6.72 & 6.66 & 6.62 & 7.72\\
        \bottomrule
    \end{tabular}
    \caption{{URA performance comparison between DL-based policies and WMMSE for $N=20$ users. Average SR is reported in $bps/Hz$. Centralized and distributed policies are considered with diverse training (where $\bm\kappa$ is randomly sampled during training) and non-diverse training (where $\bm\kappa=\bm 1$ remains constant during training) strategies. These policies are only responsible for \emph{unconstrained} RA and cannot satisfy user utility demands.}}
    \label{tab:URA}
\end{table*}

%% file: TeX/Tab/circ_gp.tex
\begin{table}[th]
    \centering
    \begin{tabular}{c c c c c}
        \toprule
        $N$ & ALCOR (cntr.) & ALCOR (dist.) & fixed DNN & GP \\
        \midrule
        10 & 5.41 & 4.37 & 6.82 & 3.86 \\
        20 & 6.56 & 6.37 & 6.73 & 3.81 \\
        50 & 6.67 & 5.95 & 6.16 & 4.73 \\
        \bottomrule
    \end{tabular}
    \caption[foo]{
        {Comparison of various RA schemes with different numbers of users and utility demands. Average SR is reported in $bps/Hz$.
        All utility demands are satisfied without violation.
        The utility demands are zero except for the following users: $N=10$: $\bm u_{1:10}^{\min}=0.25$, $N=20$: $\bm u_{1:12}^{\min}=0.1$, $N=50$: $\bm u_{1:15}^{\min}=0.05$.
        The \emph{fixed DNNs} are trained using the reward function \eqref{eq:dnn_fix} with the utility demands considered for each $N$.}
}
\label{tab:circ:GP}
\end{table}

%% file: TeX/Tab/circ_tight.tex
\begin{table}[th]
    \centering
    \captionsetup{singlelinecheck=off}
        \begin{tabular}{c c c c c c c}
            \toprule
              & \multicolumn{2}{c}{ALCOR (cntr.)} & \multicolumn{2}{c}{ALCOR (dist.)} & \multicolumn{2}{c}{fixed DNN} \\
            \cmidrule(lr){2-3} \cmidrule(lr){4-5} \cmidrule(lr){6-7}
            $N$  & SR & viol. & SR & viol. & SR & viol. \\
            \midrule
            10 & 3.13 & 0.61 & 3.85 & 0.01 & 6.35 & 37.71\\
            20 & 3.97  & 0.00  & 5.35 & 0.00 & 6.58 & 60.45 \\
            50 & 6.09  & 1.37  & 4.75  & 2.84 & 6.17 & 64.14 \\
            \bottomrule
        \end{tabular}
    \caption[foo]{
        {Stricter utility demands than the ones in \cref{tab:circ:GP}. 
        Average SR is reported in $bps/Hz$ and violation is calculated by \eqref{eq:loss}.
        The utility demand are zero except for the following users: $N=10$: $\bm u_{1:10}^{\min}=0.1$ except $u_5^{\min}=0.6, u_7^{\min}=0.8, u_9^{\min}=1$,
        $N=20$: $\bm u_{1:10}^{\min}=0.1$ except $u_5^{\min}=0.4, u_7^{\min}=0.6, u_{12}^{\min}=0.2, u_{19}^{\min}=0.8$, $N=50$: $\bm u_{20:30}^{\min}=0.3$.
        The \emph{fixed DNNs} are trained with the utility demands reported in \cref{tab:circ:GP}.}
}
\label{tab:circ:tight}
\end{table}

%% file: TeX/Tab/variousURA.tex
\begin{table*}[th] 
    \centering
    \new{
        \begin{tabular}{c c c c c}
            \toprule
            & \multicolumn{3}{c}{{ALCOR with various URA schemes}} & \multicolumn{1}{c}{ } \\
            \cmidrule(lr){2-4} 
            & centralized & GNN-based dist. & max power & RL-based RA \\
            \midrule
            average SR (bps/Hz)& 31.7  & 29.4  & 16.2  & 32.6 \\ 
            average latency (ms) & 64.1  & 77.1  & 567.8  & 55.2 \\ 
            latency violation (\%) & 1.1  & 2.3  & 90.1 & 36.1 \\ 
            convergence (ms) & 1134.9  & 962.4  & - & 4307.5 \\ 
            \bottomrule
        \end{tabular}
    }
    \caption{
        \new{Performance of ALCOR employing various URA schemes for $N=20$, compared with RL-based RA.}
    }
\label{tab:variousURA}
\end{table*}

%% file: TeX/Text/multi_channel.tex
In this section, the proposed RA scheme is assessed considering a communication system consisting of 7 cells. Each cell contains a number of transmitters and receivers, with transmitters positioned at the center and receivers uniformly distributed within the cell.
The minimum and maximum allowed distances between the transmitters and the receivers are set to $r=50$ and $R=500$ meters, respectively. 
{The carrier frequency is set to $6$ GHz, and the sub-channel bandwidth is $1$ MHz.} The noise power, $\sigma_n^2$, is set to -114 dBm, and the maximum transmit power is $p^{\max}=38$ dBm. The large-scale fading component, $\alpha_{ji}$, {which models the path loss and shadowing between transmitter $i$ and receiver $j$, is determined by $\alpha_{ji} = 120.9 + 37.6 \log_{10}(d_{ji}) + \zeta_{ji}$ dB, where $d_{ji}$ denotes the distance measured in kilometers, and $\zeta_{ji} \sim \mathcal{N}(0,\sigma_s^2)$ represents the shadowing with a standard deviation of $\sigma_s = 8$.}
The small-scale fading component follows IID circularly symmetric Gaussian distribution, \ie, $g_{ij} \sim \mathcal{CN}(0,1)$. Hence, the channel coefficients are derived as $|h_{ij}|^2=|g_{ij}|^2 / 10^{\alpha_{ij}/10}$. 
\new{The small-scale fading component changes every 10 ms and the large-scale fading component changes 10 times slower.}
Both the centralized and distributed DNNs are trained using possibly infeasible samples in the training stage. The structures of the DNNs are the same as those described in \cref{sec:sim:circ}.

\new{In this section, we assess ALCOR with a broader range of resource types and resource constraints. Hence, in addition to power allocation, the task of sub-channel assignment is added to the RA problem, where each user has access to only one sub-channel among $5$ possible sub-channels at each time instant \cite{ye_deep_2019,tian2021multiagent,ji2023meta,yuan2021meta}. Each user $i$ has a packet arrival rate following a Poisson distribution with parameter $\nu_i$, with packets waiting in a buffer queue for transmission \cite{tian2021multiagent}. Packets have $4000$ bits each. Users are either delay-sensitive, where the queue must have a maximum length, or delay-tolerant, where the queue needs only to be stable without a maximum length threshold. We consider a queue to be stable as long as its length does not increase over three consecutive batches. The average packet arrival rate of each user is randomly selected via $\nu_i \in [50, 400]$ packets per second. We consider $20$ users, among which $10$ are delay-sensitive with varying latency constraints randomly selected from $[20, 80]$ ms. For a delay-sensitive user $i$, the corresponding element in $\hat{F}(\bm H_\xi)$ in \eqref{eq:F} is defined as $q_i - d_i^{\max}$, where $q_i$ is the queue length and $d_i^{\max}$ is its threshold, depending on the user's latency constraint. For a delay-tolerant user $i$, this formulation changes to $c_i - c_i^{\max}$, where $c_i$ is a counter indicating the number of consecutive batches with increases in queue size, and $c_i^{\max} = 3$ is its threshold.
}

\new{To also study the impact of different URA schemes on ALCOR's accuracy and convergence in the mentioned communication scenario, the performance is reported while employing three different URA policies: 1) the centralized URA policy considered in the previous subsection; 2) a GNN-based distributed URA policy; and 3) a trivial URA policy (maximum power) that allocates $p^{\max}$ and a random sub-channel to active users. The architecture of the centralized URA policy is updated to have 4 layers, with the number of neurons set to $\{400, 512, 256, 120\}$, with two last layers. The first of the last two layers has 20 neurons with sigmoid activation functions for power allocation $p_i \in [0,p^{\max}], \forall i$, and the second is a categorical layer with 5 neurons and a softmax activation function for each user for sub-channel assignment.
The GNN-based distributed URA follows the principles of aggregation GNNs proposed in \cite{wang_learning_2022}. For each user policy $i$ at time instant $t$, the policy input is defined by $\bm{\mathcal{H}}_i^t \coloneqq (y_i^{1,t},\cdots,y_i^{K,t}) \in \mathbb{R}^K$, where $y_i^{k,t}$ is the $i$th element of the vector $\bm y^{k,t} \coloneqq \left[\prod_{t'=0}^{k-1} |\tilde{\bm{H}}^{{t-t'}^T}|\right] \bm 1 \in \R^N$. Here, $K=8$ defines the range of neighbors (hops) that each user considers in its input, $\tilde{\bm{H}}$ is the channel matrix $\bm{H}$ where the channel coefficients associated with the $m = 5$ most affected receivers by each user's transmitter are preserved and all other elements are set to zero, $\bm 1 \in \mathbb{R}^N$ represents a vector of all ones, and $|\cdot|$ is the pointwise absolute value operator.
A fully connected DNN with the structure $\{K, 256, 128, 1+5\}$ neurons is considered for each user, with one neuron in the last layer for power allocation and a categorical layer of 5 neurons for sub-channel assignment.
}

\new{ALCOR's performance is compared to RL-based RA, where a centralized policy is trained using REINFORCE \cite{sutton1999policy,wang_learning_2022}, with the observation state space $\mathcal{O} = \{\bm H, \bm u^{\min}, \bm q\}$, where $\bm q$ is a vector containing the queue lengths of all users. The action space $\mathcal{A}$ includes $N + 5 \times N$ values, similar to the output dimension of the centralized URA, and accounts for all users' transmit powers and sub-channels. The discounted reward $\sum_{t=0}^\infty \sum_{i=1}^N 0.9^t r_i^t$ is considered, where the reward for user $i$ at time instant $t$ is set equal to the user's data rate, $r_i^t = R_i^t$, if all the user's constraints are met, and $r_i^t = 0$ otherwise (refer to \cite{ji2023meta} for further details). Based on the state and action space dimensions mentioned above, a DNN with the structure $\{440, 512, 256, 120\}$ is used in this simulation.
}

\new{
The performance of the above-mentioned RA methods is summarized in \cref{tab:variousURA}, where ALCOR is evaluated separately with three different URA methods and compared against the RL-based RA. The comparison is made in terms of SR averaged over time, average latency, latency violation—defined as the percentage of time during which latency constraints are violated (i.e., queue lengths exceed the threshold) after convergence—and convergence rate—defined as the average time required for the method to satisfy violated constraints following any update in the latency constraints of delay-sensitive users.
According to the results, it is evident that ALCOR's performance is highly dependent on the employed URA scheme. Specifically, for the traffic load of $\nu_i \in [50, 400]$—which represents a moderate to high traffic load in the considered communication setup—the trivial maximum power URA in \cref{tab:variousURA} is unable to satisfy user latency constraints, resulting in unstable queue sizes without convergence. Queues that do not converge under this policy are excluded from the reported average latency. On the other hand, the centralized URA policy achieves high average SR and low latency while properly meeting the constraints, and it converges quickly to new constraints. The GNN-based distributed policy exhibits comparable performance. It is noted that GNN-based RA methods demonstrate higher scalability compared to the centralized approaches, thanks to the considered graph representation of the communication network \cite{wang_learning_2022}.
Although RL-based RA maintains stable queues and lower latencies with higher average SR values, it fails to satisfy all dynamic latency constraints and demonstrates a lower convergence rate. This reduced performance can be attributed to the higher complexity the policy must learn—namely, both resource allocation and constraint satisfaction. In contrast, ALCOR offloads the task of constraint satisfaction from the policy to an iterative process, allowing the policy to focus entirely on URA among the active users.
}

\new{
The employed GNN-based RA requires sharing $K \times m = 40$ variables per channel update per user. With 32-bit floating-point precision and a 10 ms channel coherence time, the communication overhead incurred by the GNN-based RA is 125 kbps per user. As mentioned in the previous subsection, the communication overhead required for sharing the normalization term is a comparably negligible value of 67 bps per user, according to the experiments.
}


%% file: TeX/Text/conclusion.tex
In this paper, a DL-based RA scheme has been proposed that utilizes DNNs as URA policies within an iterative optimization algorithm to meet dynamic user utility demands in expectation. 
The optimization algorithm addresses a time-sharing problem by optimizing the on-off status of users. In parallel, URA policies are responsible for performing unconstrained RA among active users at each time instant to maximize their SU. 
The proposed approach is agnostic to URA schemes. Consequently, depending on the employed URA policy, the proposed RA scheme can be implemented in either a centralized or distributed scenario.
Convergence analyses have been provided, relying on variational inequalities, to establish convergence guarantees.

\new{The proposed user selection procedure, based on the Bernoulli distribution, can be extended to account for user channel conditions. In this extension, users with the best channel conditions are activated while ensuring that their activation probabilities equal $\kappa_i$, as optimized by the algorithm. 
Extending ALCOR to scenarios where each user has multiple dynamic constraints would also be beneficial for many real-world applications. Additionally, exploring the application of ALCOR to other RA settings—such as computational power allocation, service migration, network slicing, and simultaneous wireless information and power transfer—under appropriate dynamic constraints, represents a compelling direction for future research.}

%% file: TeX/Text/appendix/notation.tex
In this Appendix, we present the notation, basic assumptions, and a lemma that are required for the convergence analysis.
\subsection{Notation}
The distance from $\bm\lambda \in \R^N$ to a set $\mathcal{X} \subseteq \R^N$ is given by ${\rm{dist}}(\bm\lambda, \mathcal{X}) \coloneqq \min_{\bm v \in \mathcal{X}}\nrm{\bm v - \bm\lambda}$.
An operator $\ffunc{F}{\R^N}{\R^n}$ maps each point $\bm\lambda\in\R^N$ to a set $F(\bm\lambda) \subseteq \R^n$. 
The graph of operator $F$ is denoted by ${\rm{gph}}~ F \coloneqq \{(\bm\lambda, \bm v)\in \R^{N} \times \R^n \mid \bm v \in F(\bm\lambda)\}$.
The set of zeros is defined by ${\rm{zer}}~ F \coloneqq \{\bm\lambda \in \R^N \mid 0\in F(\bm\lambda)\}$.
Operator $F$ is $\rho$-strongly monotone with some $\rho \geq 0$ if $\langle \bm v - \bm v', \bm\lambda - \bm\lambda' \rangle \geq \rho \sqnrm{\bm\lambda - \bm\lambda'}$ for all $(\bm\lambda,\bm v), (\bm\lambda', \bm v') \in {\rm{gph}}~ F$. The operator is called monotone if $\rho=0$.
The deviation between two distributions $\mathcal{D}$ and $\mathcal{D}'$ is measured using the Wasserstein-1 distance:
\begin{equation}
    W_1(\mathcal{D},\mathcal{D}') = \sup_{g \in \lip_1}\{\E[x\sim\mathcal{D}]{g(x)} - \E[y\sim\mathcal{D}']{g(y)}\},
    \label{eq:distance}
\end{equation} 
where $\lip_1$ denotes the set of $1$-Lipschitz continuous functions $\func{g}{\R^N}{\R^N}$.
During the proof, we use the Young's inequality with $a,b\in\R^n$ and $e>0$: $\sqnrm{a-b} \leq (1+e)\sqnrm{a} + (1+\nicefrac{1}{e})\sqnrm{b}$.

\subsection{Preliminary Assumptions and Lemmas}

The following assumptions are required for the convergence:
\begin{ass}[feasibility of inclusion problem \eqref{eq:timesharing}]\label{ass:feasibility}
    The inclusion problem \eqref{eq:timesharing} is feasible, namely, there exists at least one $\bm\lambda^\star\in\R^N$ such that $F(\bm\lambda^\star) \leq \bm 0$.
\end{ass}
\noindent
In \cref{alg:centralized}, ALCOR utilizes a DL model within an iterative optimization algorithm.
The DL model is integrated into the mapping $F$ within the inclusion problem \eqref{eq:timesharing}.
The feasibility of the problem then depends on several factors, including the utility demands $\bm u^{\min}$, the capability of the DL model to generalize the URA problem, the quality of training in \cref{sec:URA}, and the characteristics of the communication network, such as the quality of communication channels.
Hence, the communication network requires to have mechanisms to ensure this feasibility (refer to the supplementary material for further details).

\begin{ass}[assumption on the expected utility]\label{ass:utility}
    The expected utility $\E[\bm{H}]{\bm{U}^{\bm{\theta}^L}(\bm{H}_\xi)}$ is Lipschitz continuous in $\bm\xi$. Namely, there is a $\beta > 0$ such that for all $\bm \xi,\bm \xi'$:
    \begin{align*}
        \nrm{\E[\bm{H}]{\bm{U}^{\bm{\theta}^L}(\bm{H}_\xi)} - \E[\bm{H}]{\bm{U}^{\bm{\theta}^L}(\bm{H}_{\xi'})}}
        \leq \beta \nrm{\bm\xi - \bm\xi'}.
    \end{align*}
\end{ass}
\noindent
This assumption ensures that when the USV changes, the change in the expected utilities is bounded. 
This is easily verifiable for many utilities \eg, data rates where the resource is the transmit power.

{To continue with the convergence analysis, we recast the problem \eqref{eq:timesharing} to the following inclusion problem:
\begin{equation}
    \text{find } \bm\lambda\in\R^N ~\text{such that}~ \bm 0\in T(\bm\lambda) \coloneqq F(\bm\lambda) + A(\bm\lambda),
    \label{eq:timesharing:inclusion}
\end{equation}
with $A(\bm\lambda) \coloneqq \partial \varphi(\bm\lambda)$ as the subdifferential of the indicator function $\varphi$ defined as $\varphi(\bm\lambda) = 0$ if $\bm\lambda\in\R^N_+$ and $\varphi(\bm\lambda) = -\infty$ otherwise. 
The inclusion problem \eqref{eq:timesharing:inclusion} ensures finding a variable $\bm\lambda$ such that $F(\bm\lambda) \leq \bm 0$.
This is achieved under two conditions:
1) $\bm 0 \in T(\bm\lambda)$ with $[F(\bm\lambda)]_i=0$ and $[A(\bm\lambda)]_i=0$ for some $i\in[N]$. 
2) $\bm 0 \in T(\bm\lambda)$ with $[F(\bm\lambda)]_i < 0$ and $[A(\bm\lambda)]_i >  0$ for some $i\in[N]$. 
The second case is possible thanks to the definition of $\kappa_i$ allowing $\kappa_i$ to be nonzero when $\lambda_i = 0$, thereby permitting $[F(\bm\lambda)]_i < 0$. 
It is remarked that, by the definition of $A$, the case where $0 \in T(\bm\lambda)$ with $[F(\bm\lambda)]_i > 0$ and $[A(\bm\lambda)]_i <  0$ never occurs.}
The necessary assumptions on the mappings of inclusion problem \eqref{eq:timesharing:inclusion} are as follows:
\begin{ass}[assumptions on the mappings $F, \hat F$]\label{ass:mappings}
    \quad
    \begin{enumerate}
        \item\label{ass:mappings:1} The mapping $\func{F}{\R^N}{\R^N}$ is Lipschitz continuous with the constant $L_{{F}} \in [0,+\infty)$:\\
        $\nrm{F(\bm\lambda) - F(\bm\lambda')} \leq L_{{F}} \nrm{\bm\lambda-\bm\lambda'}, ~\forall \bm\lambda,\bm\lambda' \in\R^N_+$;
        \item\label{ass:mappings:3} Weak Minty variational inequality (WMVI) holds, \ie, there exists a nonempty set $\mathcal{S}^\star \subseteq \zer T$ such that for all $\bm\lambda^\star \in \mathcal{S}^\star$ and some $\rho \in \left(-\nicefrac{1}{2L_F},+\infty\right)$
        \[
            \langle \bm v, \bm\lambda - \bm\lambda^\star \rangle \geq \rho \sqnrm{\bm v}, \quad \text{for all} \; (\bm\lambda,\bm v) \in \graph T; 
        \]
        \vspace{-8pt}
        \item\label{ass:mappings:4} The stochastic oracle has a bounded variance: \\
        $\E[\bm{H},\bm\xi\sim \mathcal{D}^\xi(\bm\kappa)]{\|\hat F(\bm{H}_\xi) - F(\bm\lambda)\|^2} \leq \sigma^2, \forall \bm\lambda\in\R^N_+$.
    \end{enumerate}
\end{ass}
\noindent
\cref{ass:mappings:1} assumes that mapping $F$ in \eqref{eq:timesharing} and \eqref{eq:timesharing:inclusion} is Lipschitz continuous. 
This assumption can be readily fulfilled by taking \cref{ass:utility} along with the structure of distribution $\mathcal{D}^\xi$ in \cref{def:TSV}.
The following lemma investigates this assumption:
\begin{lem}[sufficient conditions for \cref{ass:mappings:1}]\label{lem:sufficientAss}
        \label{lem:sufficientAss:1} Take \cref{ass:utility}, then \cref{ass:mappings:1} holds with $L_{F} = \beta \omega$, where $\omega>0$ is the Lipschitz constant of the distribution $\mathcal{D}^\xi$ (see \eqref{eq:distance:xi} for definition).
    \begin{proof}
        This proof is closely following the proof in \cite[lem. 2.1]{drusvyatskiy_stochastic_2020} and is presented here for completeness. 
        By the Bernoulli distribution $\mathcal{D}^\xi$ associated to USVs $\bm\xi$ in \cref{def:TSV}, and also the distribution distance measure defined in \eqref{eq:distance}, the smoothness notion can be extended to the distributions: a constant $L_{\mathcal{D}^\xi} > 0$ exists such that 
        \begin{equation}
            W_1(\mathcal{D}^\xi(\bm\kappa), \mathcal{D}^\xi(\bm\kappa')) \leq L_{\mathcal{D}^\xi} \nrm{\bm\kappa - \bm\kappa'}.
            \label{eq:distance:xi}
        \end{equation}
        Moreover, as in \eqref{eq:timesharing} $\bm\kappa$ is a continuous function of $\bm\lambda$, a constant $L_\kappa > 0$ exists such that
        \begin{align*}
            \nrm{\bm\kappa(\bm\lambda) - \bm\kappa(\bm\lambda')} \leq L_\kappa \nrm{\bm\lambda - \bm\lambda'}.
        \end{align*}
        Hence, the composition is also Lipschitz continuous:
        \begin{align*}
            W_1(\mathcal{D}^\xi(\bm\kappa(\bm\lambda)), \mathcal{D}^\xi(\bm\kappa(\bm\lambda'))) \leq \omega \nrm{\bm\lambda - \bm\lambda'},
        \end{align*}
        where $\omega \coloneqq L_\kappa L_{\mathcal{D}^\xi}$ is defined.
        
        Define $g(\bm\xi) \coloneqq \langle \bm v, \E[\bm{H}]{\bm{U}^{\bm{\theta}^\star}(\bm{H}_\xi)} \rangle$ with a vector $\bm v$ where $\nrm{\bm v} \leq 1$. 
        By \cref{ass:utility}, $g(\bm\xi)$ is $\beta$-Lipschitz continuous in $\bm\xi$, since:
        \begin{align*}
            \nrm{g(\bm\xi) - g(\bm\xi')} &\leq \nrm{\bm v}\nrm{\E[\bm{H}]{\bm{U}^{\bm{\theta}^\star}(\bm{H}_\xi)}-\E[\bm{H}]{\bm{U}^{\bm{\theta}^\star}(\bm{H}_{\xi'})}}\\
            &\leq \beta \nrm{\bm\xi-\bm\xi'},
        \end{align*}
        hence:
        \begin{align*}
            \langle \bm v, F(\bm\lambda') - F(\bm\lambda) \rangle &= \E[\bm\xi\sim\mathcal{D}^\xi(\bm\kappa)]{g(\bm\xi)} - \E[\bm\xi'\sim\mathcal{D}^\xi(\bm\kappa')]{g(\bm\xi')} \\
            &\leq \beta W_1(\mathcal{D}^\xi(\bm\kappa),\mathcal{D}^\xi(\bm\kappa')),
        \end{align*}
        where in the equality the definition of $F$ in \eqref{eq:timesharing} is considered, and the inequality is due to \eqref{eq:distance} considering $g \in {\rm{lip}}_\beta$. Taking $v=\nicefrac{F(\bm\lambda') - F(\bm\lambda)}{\nrm{F(\bm\lambda') - F(\bm\lambda)}}$ and then using \eqref{eq:distance:xi} completes the proof with $L_{F} = \beta L_\kappa L_{\mathcal{D}^\xi}$.

        It is remarked that there is no control over the constants $\beta$ and $L_{\mathcal{D}^\xi}$ and they are determined by the communication network and the Bernoulli distribution, respectively. However, the constant $L_\kappa$ is adjustable by modifying $c$ in the function $\kappa_i = \nicefrac{(c+\lambda_i)}{\max_\ell\{c+\lambda_\ell\}}$ in \eqref{eq:timesharing}.
    \end{proof}
\end{lem}
\noindent
The class of nonmonotone mappings for which convergence analyses are possible is characterized by \cref{ass:mappings:3}, where nonmonotonicity can be captured thanks to the possibly negative values of $\rho$. 
The mapping $F$ defined in \eqref{eq:timesharing} may not be (strongly) monotone, since $ \langle \bm v_1 - \bm v_2, \bm\lambda_1 - \bm\lambda_2 \rangle \ngeq \rho \sqnrm{\bm\lambda_1 - \bm\lambda_2}$ with a $\rho \geq 0$ for all $(\bm\lambda_1,\bm v_1), (\bm\lambda_2,\bm v_2) \in \graph T$.
This relation can be interpreted as follows: increasing $\lambda_i$ for user $i$ results in increasing the probability of user $i$ being activated, which consequently increases its expected utility and decreases $[F(\bm\lambda)]_i$ due to the definition. 
Therefore, the relation cannot be guaranteed with a $\rho \geq 0$, whereas \cref{ass:mappings:3}, with the possibility of $\rho < 0$, readily captures the relation. 
It is worth noting that this assumption may be considered restrictive, since $\rho$ is lowerbounded by $-\nicefrac{1}{2L_F}$, a value that depends on $L_\kappa,L_{\mathcal{D}^\xi}$, and $\beta$ according to \cref{lem:sufficientAss}. Nevertheless, extensive numerical studies in \cref{sec:sim} have not witnessed divergence as long as \cref{ass:feasibility} holds (see the supplementary material where \cref{ass:feasibility} does not hold). 
Finally, \cref{ass:mappings:4} is a standard assumption in stochastic optimization literature, \eg \cite{pethick2023solving,ghadimi_mini-batch_2016}.

%% file: TeX/Text/appendix/conv.tex
With the necessary assumptions considered and the recast problem in \eqref{eq:timesharing:inclusion}, the next theorem demonstrates the convergence of the iterations generated by \cref{alg:centralized}.
\begin{thm}[subsequential convergence]\label{thm:conv}
    Suppose that \cref{ass:basic,ass:feasibility,ass:utility,ass:mappings} hold. 
    Moreover, take the stepsize sequence $\seq{\alpha_k}[k=0][K] \in (0,1)$ and $\gamma \in (0,\nicefrac{1}{L_F})$, the batch size sequence $\seq{B_k}[k=0][K] \in \N$, and suppose 
    \begin{equation*}
        \bar\mu \coloneqq 2\frac{\rho}{\gamma} + \frac{1-\sqrt{\alpha^{\max}}}{1+\sqrt{\alpha^{\max}}} - \alpha^{\max} - 2\bar\alpha \gamma^2L_{F}^2 \mathcal{A} > 0,
    \end{equation*}
    where $\alpha^{\max} \coloneqq \max_{k}\{\alpha_k\}$, $\bar\alpha\coloneqq \max_{k}\{\nicefrac{\alpha_k}{\alpha_{k+1}}\}$, and \\
    $\mathcal{A} \coloneqq 3\left(\frac{1}{\sqrt{\alpha^{\max}}(1-\gamma L_{F})^2} + \frac{1-\sqrt{\alpha^{\max}}}{\sqrt{\alpha^{\max}}}\right)$.
    Then, the iterates $\bm{\lambda}^k$ generated by \cref{alg:centralized} hold the following estimate 
    \begin{align*}
        \min_{k \in \{0,\cdots,K\}} &\E{{\rm{dist}}(\bm 0, T(\bm{\lambda}^{k}))^2} \numberthis \label{eq:conv}\\
        &\leq \frac{\sqnrm{\bar{\bm{\lambda}}^{0} - \bm{\lambda}^{\star}} + \mathcal{A} \sqnrm{\bm{h}^{-1}-\bar{\bm{\lambda}}^{-1}+\gamma F(\bar{\bm{\lambda}}^{-1})}}{\gamma^2\bar\mu \sum_{\ell=0}^K \alpha_\ell} \\
        & + \frac{\sum_{k=0}^K \left\{\alpha_k^2 Q \nicefrac{\sigma^2}{B_k} + 2\bar\alpha \alpha_k \gamma^4 L_{F}^2 \mathcal{A} \nicefrac{\sigma^2}{B_k}\right\}}{\gamma^2\bar\mu \sum_{\ell=0}^K \alpha_\ell}.
    \end{align*}
    with $Q \coloneqq \gamma^2 \mathcal{A} + \gamma^2 \alpha^{\max} (\nicefrac{\mathcal{A}}{3}+\gamma^2)$.
    \bigskip
    \begin{proof}
        Define the following operators:
        \begin{align}
            \begin{split}
                H(\bm x) &\coloneqq \bm x + \gamma F(\bm x), \\
                \bar H(\bm x) &\coloneqq \bm x + \gamma \hat{\bar{F}}(\mathbb{H}_{\bm \xi}), \quad \bm\xi\sim \mathcal{D}^\xi(\bm\kappa(\bm x)),  
            \end{split}
            \label{eq:H}
        \end{align}
        and the filtration:
        \begin{align*}
            \mathcal{F}_{k} \coloneqq &{\rm{filtration}}\{\mathbb{H}_{\bm\xi}^{0},\bar{\mathbb{H}}_{\bm\xi}^{0},\cdots,\mathbb{H}_{\bm\xi}^{k-1},\bar{\mathbb{H}}_{\bm\xi}^{k-1}\},
        \end{align*}
        where includes all the randomness involved up to iteration $k$.
        Using the notations of \cref{alg:centralized} in \eqref{eq:H}, $\bar H(\bm\lambda^k) = \bm\lambda^k + \gamma \hat{\bar{F}}(\mathbb{H}_{\bm \xi}^k)$ and $\bar H(\bar{\bm\lambda}^k) =\bar{\bm\lambda}^k + \gamma \hat{\bar{F}}(\bar{\mathbb{H}}_{\bm \xi}^k)$.
        Take the Lyapunov function \cite{pethick2023solving}:
        \begin{align*}
            \mathcal{U}_{k+1} &\coloneqq \sqnrm{\bar{\bm{\lambda}}^{k+1} - \bm{\lambda}^{\star}} \\
            &+ \mathcal{A}_{k+1}\sqnrm{\bm{h}^{k} - H(\bar{\bm{\lambda}}^k)} + \mathcal{B}_{k+1}\sqnrm{\bar{\bm{\lambda}}^{k+1} - \bar{\bm{\lambda}}^k},
            \numberthis\label{eq:lyapunov}
        \end{align*}
        with some $\mathcal{A}_{k+1}>0$ and $\mathcal{B}_{k+1}>0$ that will be defined in the sequel.
        Expanding the first term results in, 
        \begin{align*}
            \sqnrm{\bar{\bm{\lambda}}^{k+1} - \bm{\lambda}^{\star}} &= \sqnrm{\bar{\bm{\lambda}}^k - \bm{\lambda}^{\star}} \numberthis\label{eq:onestep}\\
            - 2\alpha_k &\langle \bm{h}^{k} - \bar H(\bm{\lambda}^{k}), \bar{\bm{\lambda}}^k - \bm{\lambda}^{\star} \rangle + \alpha_k^2 \sqnrm{\bm{h}^{k} - \bar H(\bm{\lambda}^{k})},
        \end{align*}
        due to \cref{alg:step:5} in \cref{alg:centralized}.
        We continue by upperbounding the terms in \eqref{eq:lyapunov} and \eqref{eq:onestep}.
        The term $\sqnrm{\bm{h}^{k} - H(\bar{\bm{\lambda}}^{k})}$ is upperbounded as follows:
        \begin{align*}
            &\bm{h}^{k} - H(\bar{\bm{\lambda}}^{k}) \label{eq:u:1} \numberthis\\
            &= \gamma\hat{\bar{F}}(\bar{\mathbb{H}}_{\bm\xi}^{k}) - \gamma F(\bar{\bm{\lambda}}^k) + (1-\alpha_k)\left(\bm{h}^{k-1} - \bar{\bm{\lambda}}^{k-1} - \gamma\hat{\bar{F}}(\bar{\mathbb{H}}_{\bm\xi}^{k})\right),\\
            &\sqnrm{\bm{h}^{k} - H(\bar{\bm{\lambda}}^{k})} 
            = (1-\alpha_k)^2 \sqnrm{\bm{h}^{k-1}-\bar{\bm{\lambda}}^{k-1}-\gamma F(\bar{\bm{\lambda}}^k)} \\
            &+\sqnrm{\alpha_k\left(\gamma F({\bar{\bm{\lambda}}^k})-\gamma\hat{\bar{F}}{(\bar{\mathbb{H}}_{\bm\xi}^{k})}\right)}\\
            &+2(1-\alpha_k) \big\langle \bm{h}^{k-1}-\bar{\bm{\lambda}}^{k-1}-\gamma F({\bar{\bm{\lambda}}^k}),
            \alpha_k(\gamma\hat{\bar{F}}{(\bar{\mathbb{H}}_{\bm\xi}^{k})}-\gamma F({\bar{\bm{\lambda}}^k})) \big\rangle,
        \end{align*}
        where the first equality is due to the definition of $H$ in \eqref{eq:H} and $\bm{h}^{k}$ in \cref{alg:centralized}.
        The second equality is also derived by adding and subtracting $(1-\alpha_k)\gamma F(\bar{\bm{\lambda}}^k)$.
        By taking the expectation conditioned on $\mathcal{F}_{k}$, the first term in the vector inner product is deterministic, while the second term vanishes due to unbiasedness of $\hat{\bar{F}}$ in \eqref{eq:F} and \eqref{eq:meanF}. Hence, the last term is zero in expectation.
        Hence,
        \begin{align*}
            &\E{\sqnrm{\bm{h}^{k} - H(\bar{\bm{\lambda}}^{k})}}[\mathcal{F}_{k}] \numberthis \label{eq:u:2}\\
            & = (1-\alpha_k)^2 \sqnrm{\bm{h}^{k-1} - \bar{\bm{\lambda}}^{k-1} - \gamma F(\bar{\bm{\lambda}}^k)}\\
            & + \alpha_k^2\gamma^2 \E{\sqnrm{\hat{\bar{F}}(\bar{\mathbb{H}}_{\bm\xi}^{k})-F(\bar{\bm{\lambda}}^k)}}[\mathcal{F}_{k}]\\
            & \leq (1+ e_k)(1-\alpha_k)^2 \sqnrm{\bm{h}^{k-1} - \bar{\bm{\lambda}}^{k-1} - \gamma F(\bar{\bm{\lambda}}^{k-1})}\\
            & + (1+\nicefrac{1}{ e_k})(1-\alpha_k)^2 \sqnrm{\gamma F(\bar{\bm{\lambda}}^k) - \gamma F(\bar{\bm{\lambda}}^{k-1})} + \gamma^2\alpha_k^2\sigma^2_k\\
            & \leq U_1 \sqnrm{\bm{h}^{k-1} - \bar{\bm{\lambda}}^{k-1} - \gamma F(\bar{\bm{\lambda}}^{k-1})} + U_2 \sqnrm{\bar{\bm{\lambda}}^k - \bar{\bm{\lambda}}^{k-1}} + U_3
        \end{align*}
        with
        \begin{align*}
            &U_1^k \coloneqq (1+ e_k)(1-\alpha_k)^2, \quad U_2^k \coloneqq (1+\nicefrac{1}{ e_k})(1-\alpha_k)^2 L_{F}^2 \gamma^2, \\
            &U_3^k \coloneqq \gamma^2\alpha_k^2\sigma^2_k, ~\text{and}~ \sigma^2_k \coloneqq \nicefrac{\sigma^2}{B_k},
        \end{align*}
        where the first inequality is derived by adding and subtracting $\gamma F(\bar{\bm{\lambda}}^{k-1})$ along with the Young's inequality with a sequence of $ e_k > 0$.
        In addition, \cref{ass:mappings:4} is invoked for a minibatch of size $B_k$ to have
        \begin{talign*}
            &\E{\sqnrm{\hat{\bar{F}}(\bar{\mathbb{H}}_{\bm\xi}^{k})-F(\bar{\bm{\lambda}}^k)}}[\mathcal{F}_{k}] \numberthis\label{eq:minibatchvar}\\
            &= \E{\sqnrm{\frac{1}{B_k} \sum_{j=1}^{B_k} \hat F(\bm{H}_\xi^j)-F(\bar{\bm{\lambda}}^k)}}[\mathcal{F}_{k}] \leq \nicefrac{\sigma^2}{B_k}.
        \end{talign*}

        To bound the last term in \eqref{eq:onestep},
        \begin{align*}
            &\alpha_k^2 \E{\sqnrm{\bm{h}^{k} - \bar H(\bm{\lambda}^{k})}}[\mathcal{F}_{k}] \numberthis\label{eq:onestep_lastterm}\\
            &= \alpha_k^2 \sqnrm{\bm{h}^{k} - H(\bm{\lambda}^{k})} + \alpha_k^2\gamma^2 \E{\sqnrm{F(\bm{\lambda}^{k}) - \hat{\bar{F}}(\mathbb{H}_{\bm\xi}^{k})}}[\mathcal{F}_{k}]\\
            &\leq \alpha_k^2 \sqnrm{\bm{h}^{k} - H(\bm{\lambda}^{k})} + \alpha_k^2\gamma^2 \sigma^2_k,
        \end{align*}
        where the first equality is by adding and subtracting $\gamma F(\bm{\lambda}^{k})$ and the fact that $\hat{\bar{F}}$ is unbiased due to \eqref{eq:F} and \eqref{eq:meanF}.
        The last inequality is also due to \cref{ass:mappings:4}.

        Similarly, the last term in \eqref{eq:lyapunov} can be bounded by
        \begin{align*}
            \E{\sqnrm{\bar{\bm{\lambda}}^{k+1} - \bar{\bm{\lambda}}^k}}[\mathcal{F}_{k}] &= \alpha_k^2 \E{\sqnrm{\bm{h}^{k} - \bar H(\bm{\lambda}^{k})}}[\mathcal{F}_{k}] \numberthis\label{eq:lyapunov_lastterm}\\
            &\leq \alpha_k^2 \sqnrm{\bm{h}^{k} - H(\bm{\lambda}^{k})} + \alpha_k^2\gamma^2 \sigma^2_k,
        \end{align*}
        due to \cref{alg:step:lambda_update} of \cref{alg:centralized} and \eqref{eq:onestep_lastterm}.

        To bound the second term in the rhs of \eqref{eq:onestep}, we refer to the following lemma:
        \begin{lem}[bounding of $-2\alpha_k\langle \bm{h}^{k} - \bar H(\bm{\lambda}^{k}), \bar{\bm{\lambda}}^k - \bm{\lambda}^{\star} \rangle$ \protect{\cite[eq. (E.7) and eq. (E.9)]{pethick2023solving}}]\label{lem:onestep_bounding}
            Take assumptions \cref{ass:mappings:1,ass:mappings:3}. Then the following bound holds:
            \begin{align*}
                -2\alpha_k&\E{\langle \bm{h}^{k} - \bar H(\bm{\lambda}^{k}), \bar{\bm{\lambda}}^k - \bm{\lambda}^{\star} \rangle}[\mathcal{F}_{k}] \\
                & \leq \alpha_k(\epsilon_1 + \frac{1}{\epsilon_2}\Delta) \E{\sqnrm{\bm{h}^{k}-H(\bar{\bm{\lambda}}^k)}}[\mathcal{F}_{k}] \\
                & -(\frac{\alpha_k}{1+\epsilon_2} \Delta + 2\alpha_k \frac{\rho}{\gamma}) \E{\sqnrm{\bm{h}^{k}-H(\bm{\lambda}^{k})}}[\mathcal{F}_{k}],
            \end{align*}
            with $\Delta \coloneqq1 - \frac{1}{\epsilon_1(1-\gamma L_{F})^2} \geq 0$, and some positive $\epsilon_1$ and $\epsilon_2$.
        \end{lem}
        \bigskip
        \noindent
        Define the following
        \begin{align}
            \begin{split}
                \mu_k & \coloneqq 2\frac{\rho}{\gamma} + \frac{\Delta}{1+\epsilon_2} - \alpha_k(1+\mathcal{B}_{k+1})\\
                X_1^k & \coloneqq \mathcal{A}_{k+1} + \alpha_k(\epsilon_1 + \frac{\Delta}{\epsilon_2})\\
                X_2^k & \coloneqq \alpha_k^2\gamma^2\sigma^2_k(1+\mathcal{B}_{k+1}).
            \end{split}
            \label{eq:X_params}
        \end{align}
        Putting bounds derived in \eqref{eq:onestep_lastterm}, \eqref{eq:lyapunov_lastterm}, and \cref{lem:onestep_bounding} into the Lyapunov function \eqref{eq:lyapunov} results in
        \begin{align*}
                \E{\mathcal{U}_{k+1}}[\mathcal{F}_{k}] &\leq \sqnrm{\bar{\bm{\lambda}}^k - \bm{\lambda}^{\star}} - \alpha_k\mu_k \E{\sqnrm{\bm{h}^{k}-H(\bm{\lambda}^{k})}}[\mathcal{F}_{k}]\\
                &+ X_1^k \E{\sqnrm{\bm{h}^{k} - H(\bar{\bm{\lambda}}^{k})}}[\mathcal{F}_{k}] + X_2^k. \numberthis
        \end{align*}
        Considering the bound for $\sqnrm{\bm{h}^{k} - H(\bar{\bm{\lambda}}^{k})}$ derived in \eqref{eq:u:2}, the following is concluded:
        \begin{align*}
                \E{\mathcal{U}_{k+1}}[\mathcal{F}_{k}] - {\mathcal{U}_{k}} \leq
                & - \alpha_k\mu_k \E{\sqnrm{\bm{h}^{k}-H(\bm{\lambda}^{k})}}[\mathcal{F}_{k}] \\
                & + \left(U_1^k X_1^k - \mathcal{A}_{k}\right) \sqnrm{\bm{h}^{k-1} - H(\bar{\bm{\lambda}}^{k-1})}\\
                & + \left(U_2^k X_1^k - \mathcal{B}_{k}\right) {\sqnrm{\bar{\bm{\lambda}}^k-\bar{\bm{\lambda}}^{k-1}}}\\
                & + U_3^k X_1^k + X_2^k, \numberthis\label{eq:descent}
        \end{align*}
        with $U_1^k, U_2^k$ and $U_3^k$ defined in \eqref{eq:u:2}.
        To guarantee descent in \eqref{eq:descent}, we need to satisfy the bounds $\epsilon_1 >0, \epsilon_2 > 0$, and $\Delta \geq 0$ defined in \cref{lem:onestep_bounding}, along with $\mu_k > 0$, $U_1^k X_1^k - \mathcal{A}_{k} \leq 0$, and $U_2^k X_1^k - \mathcal{B}_{k} \leq 0$. Hence, with $\alpha^{\max} \coloneqq \max_{k}\{\alpha_k\}$, the parameters are set as follows:
        \begin{align*}
            \bullet~&\Delta \geq 0 ~\implies~ \epsilon_1 \geq \nicefrac{1}{(1-\gamma L_{F})^2}, ~\text{set:}~ \epsilon_1=\nicefrac{1}{\sqrt{\alpha^{\max}}(1-\gamma L_{F})^2}\\
            &\implies \Delta=1-\sqrt{\alpha^{\max}}\\
            \bullet~&U_1^k X_1^k - \mathcal{A}_{k} \leq 0 ~\longrightarrow~ ~\text{take:}~ \mathcal{A}_k=\mathcal{A} ~\longrightarrow~\\
            &(1-\alpha_k)^2(1+e_k) \mathcal{A} + (1-\alpha_k)^2(1+e_k) \alpha_k (\epsilon_1 + \frac{\Delta}{\epsilon_2}) - \mathcal{A} \\
            &\leq -\alpha_k \mathcal{A} + e_k(1-\alpha_k)\mathcal{A} + (1-\alpha_k)^2(1+e_k) \alpha_k (\epsilon_1 + \frac{\Delta}{\epsilon_2}) \\
            &\leq -\alpha_k (1-\nicefrac{e_k}{\alpha_k})\mathcal{A} + (1+e_k) \alpha_k (\epsilon_1 + \frac{\Delta}{\epsilon_2}) \leq 0\\
            &\implies~ \mathcal{A}\geq \frac{(1+\nu)(\epsilon_1+\nicefrac{\Delta}{\epsilon_2})}{1-\nu}, ~\text{with}~\nu\coloneqq \nicefrac{e_k}{\alpha_k} \in (0,1),\\
            &\text{set:}~ \epsilon_2=\sqrt{\alpha^{\max}} \implies \\
            &\mathcal{A} = \frac{1+\nu}{1-\nu}\left(\frac{1}{\sqrt{\alpha^{\max}}(1-\gamma L_{F})^2} + \frac{1-\sqrt{\alpha^{\max}}}{\sqrt{\alpha^{\max}}}\right)\\
            \bullet~&U_2^k X_1^k - \mathcal{B}_{k} \leq 0 ~\implies~ ~\text{take:}~ \mathcal{B}_{k}=\frac{U_2^k}{U_1^k}\mathcal{A} = \frac{1}{e_k}\gamma^2L_{F}^2 \mathcal{A}\\
            \bullet~&\mu_k > 0 ~\longrightarrow~ \mu_k=2\frac{\rho}{\gamma} + \frac{\Delta}{1+\epsilon_2} - \alpha_k(1+\mathcal{B}_{k+1}) \numberthis\label{eq:barmu}\\
            & \quad\quad\quad = 2\frac{\rho}{\gamma} + \frac{\Delta}{1+\epsilon_2} - \alpha_k - \frac{\alpha_k}{e_{k+1}} \gamma^2L_{F}^2 \mathcal{A} \\
            & \quad\quad\quad \geq 2\frac{\rho}{\gamma} + \frac{1-\sqrt{\alpha^{\max}}}{1+\sqrt{\alpha^{\max}}} - \alpha^{\max} - \frac{\bar\alpha}{\nu} \gamma^2L_{F}^2 \mathcal{A} \eqqcolon \bar\mu \geq 0\\
            & \quad\quad\quad~\text{with}~ \bar\alpha\coloneqq \max_{k}\{\nicefrac{\alpha_k}{\alpha_{k+1}}\},
        \end{align*}
        where in the first bullet, $\epsilon_1$ is chosen due to $\alpha^{\max} \in (0,1)$.
        In the second bullet, the first inequality is due to $(1-\alpha_k)^2 < 1-\alpha_k$, and the second inequality is due to $(1-\alpha_k)^2 < 1$, and $(1-\alpha_k) < 1$.
        The third bullet uses the relation of $U_2^k X_1^k - \mathcal{B}_{k}$ with $U_1^k X_1^k - \mathcal{A}_{k}$.
        In the last bullet, $\alpha^{\max} \geq \alpha_k$ is considered due to the diminishing stepsize assumption.
        The last term in \eqref{eq:descent} can also be upperbounded as
        \begin{align*}
            U_3^k X_1^k + X_2^k \leq \alpha_k^2 \sigma^2_k Q + \bar{\alpha} \alpha_k \sigma^2_k \gamma^4 \nicefrac{L_{F}^2 \mathcal{A}}{\nu},
        \end{align*}
        where $Q \coloneqq \gamma^2 \mathcal{A} + \gamma^2 \alpha^{\max} (\epsilon_1+\nicefrac{\Delta}{\epsilon_2})+\gamma^2$, $\alpha^{\max} \geq \alpha_k$ and the definition of $\nu$ is considered.
        Putting the above inequalities back into \eqref{eq:descent} results in 
        \begin{align*}
            \E{\mathcal{U}_{k+1}}[\mathcal{F}_{k}] - {\mathcal{U}_{k}} \leq
                & - \alpha_k\mu_k \E{\sqnrm{\bm{h}^{k}-H(\bm{\lambda}^{k})}}[\mathcal{F}_{k}] \\
                & + \alpha_k^2 \sigma^2_k Q + \bar{\alpha} \alpha_k \sigma^2_k \gamma^4 \nicefrac{L_{F}^2 \mathcal{A}}{\nu}. \numberthis\label{eq:descent:final}
        \end{align*}
        Taking the total expectation, rearranging, and telescoping results in
        \begin{align*}
            \bar\mu \sum_{k=0}^K  \alpha_k &\E{\sqnrm{\bm{h}^{k}-H(\bm{\lambda}^{k})}} \leq \mathcal{U}_{0} - \E{\mathcal{U}_{K+1}} \\
            &+ \textstyle \sum_{k=0}^K \{\alpha_k^2 \sigma^2_k Q + \bar{\alpha} \alpha_k \sigma^2_k \gamma^4 \nicefrac{L_{F}^2 \mathcal{A}}{\nu}\}
        \end{align*}
        where $\mu_k$ is lowerbounded by $\bar\mu$, defined in \eqref{eq:barmu}. Omitting $\E{\mathcal{U}_{K+1}} > 0$ from rhs and dividing both sides by $\bar\mu \sum_{k=0}^K \alpha_k$ gives
        \begin{talign*}
            &\sum_{k=0}^K  \frac{\alpha_k}{\sum_{\ell=0}^K \alpha_\ell} \E{\sqnrm{\bm{h}^{k}-H(\bm{\lambda}^{k})}} \numberthis\label{eq:descent:proof} \\
            &\leq \frac{\sqnrm{\bar{\bm{\lambda}}^{0} - \bm{\lambda}^{\star}} + \mathcal{A}\sqnrm{\bm{h}^{-1}-\bar{\bm{\lambda}}^{-1}+\gamma F(\bar{\bm{\lambda}}^{-1})}}{\bar\mu \sum_{\ell=0}^K \alpha_\ell} \\
            & + \frac{\sum_{k=0}^K \left\{\alpha_k^2 \sigma^2_k Q + \bar{\alpha} \alpha_k \sigma^2_k \gamma^4 \nicefrac{L_{F}^2 \mathcal{A}}{\nu}\right\}}{\bar\mu \sum_{\ell=0}^K \alpha_\ell},
        \end{talign*}
        where initializations $\bar{\bm\lambda}^{-1}=\bar{\bm\lambda}^0$ are considered in $\mathcal{U}_0$. Moreover, 
        \begin{align*}
            \sqnrm{\bm{h}^{k}-H(\bm{\lambda}^{k})} = \sqnrm{\gamma T(\bm{\lambda}^{k})} \geq {\rm{dist}}(\bm 0, \gamma T(\bm{\lambda}^{k}))^2,
        \end{align*}
        where $\bm{h}^{k} \in \bm{\lambda}^{k} + \gamma A(\bm{\lambda}^{k})$ is used due to \cref{alg:step:barz} of \cref{alg:centralized}, as the step imposes the update $\bm{\lambda}^{k} = (\id + \gamma A)^{-1} \bm{h}^{k} = \max\{\bm 0, \bm{h}^{k}\}$, where $\id$ is the identity function and $(\id + \gamma A)^{-1}$ is the inverse of the mapping $\id + \gamma A$ which solves $\bm{\lambda}^{k} = \argmin_{\bm u}\left\{\varphi(\bm u)+\nicefrac{1}{2\gamma}\sqnrm{\bm u-\bm h^k}\right\}$ with $\varphi$ defined in \eqref{eq:timesharing:inclusion}.
        The lhs of \eqref{eq:descent:proof} can also be lowerbounded by the $\min\{\cdot\}$ operator as it is a weighted sum.
        Hence, 
        \begin{align*}
            \min_{k \in \{0,\cdots,K\}} &\E{{\rm{dist}}(\bm 0, T(\bm{\lambda}^{k}))^2}\\
            &\leq \frac{\sqnrm{\bar{\bm{\lambda}}^{0} - \bm{\lambda}^{\star}} + \mathcal{A}\sqnrm{\bm{h}^{-1}-\bar{\bm{\lambda}}^{-1}+\gamma F(\bar{\bm{\lambda}}^{-1})}}{\gamma^2\bar\mu \sum_{\ell=0}^K \alpha_\ell} \\
            & + \frac{\sum_{k=0}^K \left\{\alpha_k^2 \sigma^2_k Q + \bar{\alpha} \alpha_k \sigma^2_k \gamma^4 \nicefrac{L_{F}^2 \mathcal{A}}{\nu}\right\}}{\gamma^2\bar\mu \sum_{\ell=0}^K \alpha_\ell}.
        \end{align*}
        Setting $\nu = 0.5$, and using \eqref{eq:minibatchvar} completes the proof.
    \end{proof}
\end{thm}
\noindent
Following the extragradient-like updates as in \cref{alg:centralized}, the same estimate for $\E{{\rm{dist}}(\bm 0, T(\bm{\lambda}^{k}))^2}$---indicating $\bm 0\in T(\bm{\lambda}^{k})$ in \eqref{eq:timesharing:inclusion} for the class of nonmonotone mappings---has been presented in \cite{pethick2023solving}, except for the last term on the rhs of \eqref{eq:conv} with $\alpha_k$ in the numerator rather than $\alpha_k^2$, indicating that an increasing batch size is necessary. This difference is due to the independence of the stochastic oracle $\hat F$ from the decision variables $\bm\lambda$ and $\bar{\bm\lambda}$.
Moreover, the condition type on $\bar\mu$ in \cref{thm:conv} can also be found in \cite{pethick2023solving}, and it can be controlled by adjusting $\alpha^{\max}$ as well as $L_{F}$.
It should be mentioned that the Lipschitz constant $L_{F}$ can be chosen arbitrarily small (see \cref{lem:sufficientAss}).

The estimate in \eqref{eq:conv} indicates the possibility of achieving a diminishing rhs, and consequently, convergence to a fixed point, by selecting suitable stepsizes $\alpha_k$ and batch sizes $B_k$.  
The following remark highlights the convergence rate achieved through the standard choice of these parameters \cite{ghadimi_mini-batch_2016}:
\begin{rem}\label{rem:rate}
    Consider either of the following two scenarios, with appropriate stepsize $\alpha_k$ and Lipschitz constant $L_{F}$ (see \cref{lem:sufficientAss}) that satisfies $\bar{\mu} > 0$:
    \begin{itemize}
        [leftmargin=10pt,]
        \item Take fixed stepsize $\alpha_k = \alpha$ and fixed batch size $B_k=\sqrt{K}$;
        \item Take diminishing stepsize $\alpha_k = \frac{\alpha_0}{\sqrt{1 + \tilde{\alpha}k}}$ and the increasing batch size $B_k=1+\sqrt{k}$, with some positive $\alpha_0$ and $\tilde{\alpha}$. 
    \end{itemize}
    Then, the convergence rate is 
    \[
        \min_{k \in \{0,\cdots,K\}} \E{{\rm{dist}}(0, T(\bm{\lambda}^{k}))^2} \leq \mathcal{O}(\nicefrac{1}{\sqrt{K}}).
    \]
\end{rem}